\documentclass[journal]{IEEEtran}
\IEEEoverridecommandlockouts

\usepackage{amsthm,amssymb,graphicx,multirow,amsmath,color,amsfonts}%,ulem}
\usepackage[update,prepend]{epstopdf}
\usepackage[noadjust]{cite}
\usepackage[latin1]{inputenc}
\usepackage{tikz}
\usetikzlibrary{arrows,calc}		% Optional ticks libraries
\usepackage{bbm} % for \mathbbm{1}
\usepackage{pdfpages}
\usepackage{tabulary}
\usepackage{multirow}
\usepackage{comment}
\usepackage{subcaption} % For subfigures
\usepackage{caption}  % For customizing captions
\usepackage[ruled,vlined,linesnumbered]{algorithm2e}
\pdfoutput=1
\usepackage{svg}
\usepackage{enumitem,kantlipsum}
\usepackage{graphicx} % Required for resizing
\def\chb#1{{\color{blue}#1}}

\renewcommand{\chb}[1]{\textcolor{black}{#1}}

\usepackage{tabularx}
\usepackage{makecell}               % For wrapping text in cells (especially headers)
\usepackage{array}                  % For extended column definitions
\usepackage{balance}
\usepackage{cuted}
\usepackage{arydshln}
\usepackage{afterpage}
\everymath{\footnotesize} % Applies \small to inline math
\usepackage{xcolor}
\usepackage{tcolorbox} % More flexible than fcolorbox

\def\nb0{{\mathbf{0}}}
\def\nb1{{\mathbf{1}}}

\newtheorem{lemma}{Lemma}

\newtheorem{ndef}{Definition}

\newtheorem{prop}{Proposition}

\newtheorem{remark}{Remark}

\newtheorem{conjecture}{Conjecture}
	
\allowdisplaybreaks % Allows breaking of eqnarray over multiple pages (avoids unnecessary blanks in the document before eqnarray)

\begin{document}
\graphicspath{{./Figures/}}
\title{
% Generic Construction of Arbitrary-Length CAZAC Sequences and Application of Bj\"orck Sequences to Wireless Systems
{%\fontsize{20.8}{19}\selectfont 
\huge Bj\"orck Sequences: Extension to Arbitrary Lengths, Correlation Analysis, and Applications to Wireless Systems}
}
\author{
Harish K. Dureppagari, Chiranjib Saha, R. Michael Buehrer, Harpreet S. Dhillon
\thanks{H. K. Dureppagari, R. M. Buehrer, and H. S. Dhillon are with Wireless@VT, Department of ECE, Virginia Tech, Blacksburg, VA 24061, USA. Email: \{harishkumard, rbuehrer, hdhillon\}@vt.edu. C. Saha is with Qualcomm Technologies Inc., San Diego, CA 92121, USA. Email: csaha@qti.qualcomm.com. The support of the US NSF (Grant CNS-2107276) is gratefully acknowledged. This paper has been published in part at the IEEE MILCOM 2025, Los Angeles, CA, USA~\cite{11310769}.
}
\vspace{-8mm}
}

\maketitle

\begin{abstract}
In this paper, we propose a sequence construction framework that extends prime-length Bj\"orck sequences, a class of Constant Amplitude Zero Autocorrelation (CAZAC) sequences, to arbitrary lengths using Goldbach's conjecture for even and odd integers. The framework is generic and applies to any CAZAC family defined for prime lengths and supports extensions to both cyclically shifted sequences and sequences with different root indices. We analytically characterize the resulting correlation behavior and show that the construction preserves orthogonality among cyclic shifts while maintaining favorable zero-lag cross-correlation across different root-index sequences. We further investigate Bj\"orck sequences as candidates for reference signals in next-generation wireless systems. Using the proposed framework, we extend Bj\"orck sequences to arbitrary lengths and evaluate their time- and frequency-offset estimation performance in terrestrial (TNs) and non-terrestrial networks (NTNs). Results show performance comparable to Zadoff--Chu (ZC) sequences in low-Doppler TN environments and improved robustness in high-Doppler NTN scenarios due to superior ambiguity-function properties. We also identify an inherent Doppler-dependent behavior that can cause sequence misidentification under large Doppler shifts. To address this, we propose two mitigation strategies: (i) leveraging coarse Doppler estimates prior to detection, and (ii) selecting appropriately spaced subsets of orthogonal sequences. Ambiguity function-based analysis demonstrates the effectiveness of these approaches in improving estimation reliability. Overall, this work enables practical arbitrary-length CAZAC sequence design and establishes Bj\"orck sequences as a strong alternative for reference signal design in high-Doppler environments.
\end{abstract}

\begin{IEEEkeywords}
Generic Sequence Construction, inner-product, Bj\"orck sequences, CAZAC Sequences, reference signals, periodic cross-correlation, aperiodic cross-correlation, ambiguity function, TN, NTN.
\end{IEEEkeywords}
\section{Introduction}\label{sec:intro}

Sequence design is a fundamental aspect of wireless systems, with the choice of sequences tailored to specific applications and performance requirements~\cite{tse2005fundamentals,144727}. In modern wireless systems, CAZAC sequences have emerged as an attractive class of sequences due to their ideal correlation properties and constant envelope characteristics. In particular, ZC sequences~\cite{1054840,andrews2023primerzadoffchusequences} are widely adopted in Long Term Evolution (LTE), and 5G New Radio (NR) systems for synchronization and uplink random access~\cite{3gpp::38211}. Looking ahead to 6G, ongoing discussions on Release 20 study items indicate a renewed focus on sequence design across multiple functionalities, including uplink (UL) random access~\cite{R1_RA_FLsummary}, UL channel state information (CSI) acquisition~\cite{R1_UL_CSI_FLsummary}, and downlink (DL) CSI acquisition~\cite{R1_DL_CSI_FLsummary}. While existing sequences used in 5G NR (e.g., ZC for UL random access) are retained as the baseline solutions, there is a growing consensus to explore alternative sequence families that can better address emerging requirements such as high mobility and operation in NTNs. In particular, low earth orbit (LEO)-based NTN systems are characterized by large propagation delays and Doppler differences across satellites and rapidly varying Doppler shifts~\cite{10542356,dureppagari_ntn_10355106,dureppagari2024leo,dureppagari2026leobasedcarrierphasepositioning6g}. Under such conditions, the correlation properties of ZC sequences degrade as Doppler-induced distortions alter their phase structure, leading to spurious correlation peaks and degraded delay and Doppler estimation performance. Furthermore, signals transmitted from multiple satellites may arrive at the user equipment (UE) with partial temporal overlap, making aperiodic cross-correlation behavior more critical than periodic correlation in practical NTN scenarios. Although NTN integration into 5G and beyond includes integrating LEO, medium earth orbit (MEO), and geostationary earth orbit (GEO) systems into TN systems, we focus on LEO-based NTN systems in this work due to their superior link budgets, lower deployment costs, and the increasing commercial interest in large-scale LEO constellation deployments~\cite{dureppagari_ntn_10355106,dureppagari2024leo}.

These challenges motivate the need for sequence designs that are inherently robust to high-Doppler and partial-overlap conditions. In this work, we explore Bj\"orck sequences~\cite{Bjorck1990,bjorck1995new,Saffari2001}, a class of CAZAC sequences with favorable ambiguity function properties, as a promising candidate for reference signal design in next-generation wireless systems. Bj\"orck sequences exhibit improved delay and Doppler estimation performance under large Doppler shifts, making them particularly well suited for NTN environments. A fundamental limitation in practical deployments, however, lies in extending CAZAC sequences beyond their inherent prime-length definitions. In current 4G or 5G systems, prime-length sequences are typically adapted to non-prime lengths using repetition-based approaches~\cite{3gpp::38211}. While simple, these methods disrupt the underlying sequence structure, leading to loss of orthogonality among cyclically shifted versions and increased zero-lag cross-correlation (inner product) across sequences with different root indices. More broadly, existing analytical constructions of CAZAC sequences remain restricted to specific length classes, while numerical or algebraic extensions either lack structural guarantees or depend on stringent combinatorial conditions, limiting their applicability to arbitrary lengths~\cite{zhang2023doppler, correll2025rootofunitycazac, MowsUnifiedConstruction, amis2025cazac}. Motivated by these limitations, we propose a unified sequence construction framework for extending prime-length CAZAC sequences to arbitrary lengths. The proposed approach preserves key correlation properties, including orthogonality among cyclic shifts and controlled cross-correlation across sequence sets, while providing flexibility in sequence design. This makes the resulting sequences well-suited for both TN and NTN deployments.

\subsection{Prior Work}\label{sec:prior_work}
In the context of our contributions, we discuss two main lines of prior work: (i) Bj\"orck sequences and their properties, and (ii) sequence construction methods, particularly those aimed at extending prime-length CAZAC sequences to non-prime or arbitrary lengths.

{\em Prior work on Bj\"orck sequences.} 
Bj\"orck sequences were first introduced in~\cite{Bjorck1990}, with detailed constructions and classifications further developed in~\cite{bjorck1995new,Saffari2001}. A framework for generating orthogonal sets of Bj\"orck sequences was presented in~\cite{BjrckSequenceSets}, while their periodic and aperiodic ambiguity properties, along with those of related CAZAC families such as Wiener sequences, have been analyzed in~\cite{BjrckAmbiguity4775877,4250292}. Their Fourier dual properties were studied in~\cite{fourierdualsbjorck}. In addition, Bj\"orck sequences have been explored in radar applications due to their favorable ambiguity characteristics~\cite{4339445}, and in cellular systems to enhance physical random access channel (PRACH) capacity~\cite{6089366}. Despite these studies, their potential as reference signals in modern wireless systems, particularly under high-Doppler conditions such as NTN, has not been systematically investigated.

{\em Prior work on non-prime length sequence construction.} 
%In practical wireless systems such as LTE and NR, prime-length CAZAC sequences are often extended to non-prime lengths using repetition- or truncation-based approaches~\cite{3gpp::38211}. While simple to implement, these methods degrade correlation properties by introducing nonzero inner products between cyclically shifted sequences and increasing zero-lag cross-correlation between different sequence sets.
Recent efforts have focused on optimizing parameters of existing CAZAC sequences, such as root index selection for improved Doppler resilience~\cite{zhang2023doppler}, or performing exhaustive searches over root-of-unity CAZAC sequences~\cite{correll2025rootofunitycazac}. These approaches largely rely on Mow's unified quadratic construction~\cite{MowsUnifiedConstruction}, which inherently does not support arbitrary lengths. While numerical approaches have also been proposed to generate near-CAZAC sequences of arbitrary lengths using iterative projection methods~\cite{amis2025cazac}, they lack analytical structure, reproducibility, and control over correlation properties. Furthermore, recent works have extended Bj\"orck-type constructions using algebraic frameworks such as group rings and partial difference sets~\cite{arasu2023bjorck}, yielding nearly perfect sequences (small autocorrelation), but they do not support arbitrary lengths.

Despite this extensive body of work, two key gaps remain: 
1) Bj\"orck sequences have not been explored as a candidate for reference signals for the next-generation wireless systems; 
2) there is no unified analytical framework for extending prime-length CAZAC sequences to non-prime lengths while preserving desirable correlation and inner-product properties. 

In this work, we bridge these gaps by: 
1) investigating Bj\"orck sequences as a potential candidate for reference signals, highlighting their advantages, particularly in high-Doppler environments such as NTN; and 
2) proposing a unified sequence construction framework that extends prime-length CAZAC sequences to non-prime lengths for both cyclic-shift and different root-index sequences, along with a comprehensive analysis of their correlation properties.

\subsection{Contributions}\label{sec:contributions}
The main contributions of this paper are outlined as follows.
\begin{itemize}[wide, labelwidth=!, labelindent=0pt]

    \item {\em \textbf{Generic Sequence Construction Method:}} 
    We propose a generic sequence construction framework for extending CAZAC sequences, which are inherently defined for prime lengths, to arbitrary lengths. Specifically, we leverage Goldbach's conjecture for both even and odd integers to construct arbitrary-length sequences by extending: (i) cyclically shifted versions, and (ii) sequences corresponding to different root indices. To quantify the impact of the proposed construction, we analyze the inner-product behavior and show that it preserves orthogonality under cyclic shifts and maintains favorable zero-lag cross-correlation properties across different root indices. Furthermore, we demonstrate that the proposed framework enables a flexible trade-off between the total number of available sequences and the maximum number of orthogonal sequences, which can be exploited to improve interference resilience.

    \item {\em \textbf{Applicability of Bj\"orck Sequences to Wireless:}} 
    We investigate Bj\"orck sequences as a potential candidate for reference signals in next-generation wireless systems. In particular, we employ the proposed construction to extend cyclically shifted prime-length Bj\"orck sequences to arbitrary lengths. To evaluate their applicability, we study the time- and frequency-offset estimation performance of these sequences in comparison with widely used CAZAC sequences, such as ZC. Specifically, we evaluate two scenarios: (i) low-Doppler environments representing TN systems, and (ii) high-Doppler environments representing NTN systems. Our results demonstrate that Bj\"orck sequences are a promising alternative class of CAZAC sequences for a reference signal candidate and offer a potential replacement for ZC sequences in next-generation wireless systems, particularly in high-Doppler NTN scenarios.

    \item {\em \textbf{Doppler-Dependent Behavior and Mitigation Approaches:}} 
    While Bj\"orck sequences provide accurate delay and Doppler estimation in high-Doppler environments, we identify an inherent Doppler-dependent behavior that can lead to sequence misidentification under large Doppler shifts, thereby degrading estimation accuracy. To address this challenge, we propose two mitigation approaches: (1) leveraging a coarse Doppler estimate prior to reference signal detection to improve detection accuracy and (2) selecting a subset of available orthogonal sequences to maintain a minimum separation between sequences to account for the maximum Doppler uncertainty. Using ambiguity function-based analysis, we illustrate the Doppler-dependent behavior and demonstrate the effectiveness of the proposed approaches in mitigating misidentification. %Lastly, to support large LEO constellations serving numerous UEs across different regions, we propose sequence reuse strategies that maximize the efficient utilization of available sequences, enabling scalability and broader coverage for large-scale NTN deployments.

\end{itemize}

{\em Notations.} The set of integers is denoted by \(\mathbb{Z}\). For two integers \(a\) and \(b\), the notation \(a \equiv b \pmod{n}\) indicates that \(a\) is congruent to \(b\) modulo \(n\), where \(n\) is a positive integer. Conversely, \(a \not\equiv b \pmod{n}\) denotes \(a\) is not congruent to \(b\) modulo \(n\). \(a\) is a quadratic residue modulo \(n\) if \(a \equiv x^2 \pmod{n}\), for some \(x\in\mathbb{Z}, x\not\equiv 0\pmod{n}\). On the other hand, \(a\) is a quadratic nonresidue modulo \(n\) if \(a \not\equiv x^2 \pmod{n}\), for any $x\in\mathbb{Z}$. \(\genfrac{(}{)}{}{}{a}{b}\) denotes the Legendre symbol. %{The normalized correlation between complex vectors \(\mathbf{x}\) and \(\mathbf{y}\) is denoted by \(\frac{\mathbf{x}^H\mathbf{y}}{(\mathbf{x}^H\mathbf{x})(\mathbf{y}^H\mathbf{y})}\), where \(H\) denotes conjugate transpose.}

\section{Generic Framework for Extending Prime-Length CAZAC Sequences to Non-Prime Lengths}\label{sec:generic_construction_cazac}
In this section, we propose a generic construction to extend prime-length CAZAC sequences to generate non-prime-length sequences. Such a construction is particularly important for practical wireless systems, such as OFDM systems, where sequence lengths are typically not prime. Consequently, when CAZAC sequences, which are primarily defined for prime lengths in their original form~\cite{andrews2023primerzadoffchusequences,Bjorck1990,bjorck1995new}, are employed in such systems, they must be adapted to support non-prime sequence lengths. CAZAC sequences are typically generated from a base sequence, from which additional orthogonal sequences can be obtained through cyclic shifts and, in certain cases, through different root indices. For a base sequence of length \(Q\), where \(Q\) is a prime number, one can generate \(Q\) cyclically shifted versions which are mutually orthogonal. For certain CAZAC families, such as ZC sequences, additional sequences can be obtained by varying root indices. Specifically, one can generate \(Q-1\) distinct base sequences corresponding to root indices in the range \([1,\,Q-1]\). For each root sequence, \(Q\) cyclically shifted versions can be obtained, resulting in a total of \(Q\cdot(Q-1)\) distinct sequences. In OFDM systems, the sequence length is determined by the number of subcarriers, denoted by \( N \), which depends on the available bandwidth and subcarrier spacing (SCS). Typically, \( N \) is an even number and a multiple of \(12\). In such a case, for generating a sequence of length \( N \), denoted by \(\mathbf{s}_N\), the largest prime number \( Q \) less than \( N \) is selected, base sequence of length \(Q\) is generated, denoted by \(\mathbf{s}_Q\). After that, to generate an \( N \)-length sequence, the traditional approach involves repeating the samples of a given base sequence as follows
\begin{align}
    s_N[m] = s_Q[m \bmod Q], \quad m = 0,1,\cdots,N-1.
    \label{eq:non_prime_seq_rep}
\end{align}

Note that, using cyclic shifts of the base sequence \(\mathbf{s}_Q(\cdot)\), one can generate \(Q\) mutually orthogonal sequences of length \(Q\) for each root index, each of which can then be extended to support the desired non-prime length \(N\) using~\eqref{eq:non_prime_seq_rep}. A similar approach is used for ZC sequences in 5G NR~\cite{3gpp::38211} for preamble generation as part of PRACH transmission. In this context, cyclically shifted versions of a root sequence are primarily employed to generate multiple orthogonal preambles, while different root indices are utilized only when the number of required preambles exceeds the capacity provided by allowed cyclic shifts of a single root sequence.

Having said that, the repetition-based extension in~\eqref{eq:non_prime_seq_rep} alters the orthogonality property among cyclically shifted versions of the base sequence. Specifically, after extending the sequences to a non-prime length \(N\), the inner product (equivalently, the zero-lag cross-correlation) between two cyclically shifted versions of the base sequence is no longer zero, and the degree of non-orthogonality increases as the difference between \( Q \) and \( N \) becomes larger. To illustrate this effect, we consider 
% a representative example. First, consider the case where the desired sequence length is \(N = 60\). The largest prime number less than \(60\) is \(Q = 59\). In this case, a base sequence of length \(59\) is generated, and the first sample is repeated to obtain a sequence of length \(60\). The cyclically shifted versions of the prime-length base sequence can then be extended in the same manner to obtain multiple sequences of length \(60\). Since the difference between \(N\) and \(Q\) is only one sample, the inner product between any two extended cyclically shifted sequences becomes \(\frac{1}{N} = \frac{1}{60}\), rather than zero as in the original prime-length construction. Next, consider
a sequence length \(N = 120\). The largest prime number less than \(120\) is \(Q = 113\). Thus, a sequence of length \(113\) is generated and the first \(7\) samples are repeated to extend the sequence to length \(120\). In this case, the zero-lag cross-correlation between any two extended cyclically shifted sequences is empirically observed to be uniformly distributed between \(\frac{1}{120}\) and \(\frac{7}{120}\), yielding an average inner product of approximately \(\frac{4}{120}\). This increase in the zero-lag cross-correlation is undesirable in practical systems. In particular, when cyclically shifted versions of the base sequence are assigned to different interfering transmitters and their corresponding signals arrive with significant or complete overlap at the receiver--as is often the case in TNs due to negligible delay and Doppler differences among interfering signals--the resulting non-zero inner product can lead to increased interference and degraded detection performance.

To address the non-orthogonality issue discussed above and to introduce greater flexibility in constructing sequences for arbitrary lengths, we propose an alternative framework for extending prime-length CAZAC sequences to non-prime lengths. The proposed construction ensures that the inner product between cyclically shifted versions of the extended sequences is either zero or a constant value for any arbitrary length \(N\), regardless of the difference between the sequence length and the largest prime number less than that. Furthermore, the proposed construction provides flexibility in selecting the number of orthogonal sequences that can be generated, even after extension. %This enables assigning mutually orthogonal sequences for desired transmissions and interfering signals, thereby providing better control over the resulting interference levels. 
Importantly, the proposed framework is general and applicable to any CAZAC sequence family that is inherently defined for prime lengths. Specifically, the proposed approach leverages Goldbach's conjecture~\cite{helfgott_goldbach_sw,estermann_goldbach_s,heathbrown_goldbach_s,helfgott_goldbach_2_sw,Saouter_goldbach_w}, which, although not fully proven, has been numerically validated up to very large limits, making it practically applicable to engineering problems. Goldbach's conjecture for even and odd integers is defined as follows.
%Furthermore, for prime-length CAZAC sequences, zero-lag cross-correlation between sequences correspond to different roots, e.g., ZC sequences, is constant which is not the case with the new construction, but it does not substantially different as it uniformly
\begin{conjecture}[Binary or Strong Goldbach's Conjecture]
Every even integer \(n\) greater than two can be expressed as the sum of two primes.
\end{conjecture}
\begin{conjecture}[Ternary or Weak Goldbach's Conjecture]
Every odd integer \(n\) greater than five can be expressed as the sum of three primes.
\end{conjecture}

With this background, we now present a generic sequence construction framework for both even and odd integers. Although the proposed construction can be applied to extend sequences generated with different cyclic shifts or root indices, the primary objective is to preserve orthogonality among a subset of cyclically shifted versions of a base sequence. At the same time, the construction ensures that sequences generated with different root indices exhibit favorable inner-product behavior.

% \subsection{Generic Construction: Even-Length Sequences}\label{sec:even_len_bjrck_seqs}
When \(N\) is an even integer, Goldbach's conjecture guarantees that it can be expressed as the sum of two prime numbers, i.e.,
\begin{align}
    N = Q_1 + Q_2,
    \label{eq:goldbach_even_generic}
\end{align}
where both \(Q_1\) and \(Q_2\) are prime. Without loss of generality, we assume \(Q_1=\max(Q_1,Q_2)\) to maintain consistency in the construction. To generate an \(N\)-length sequence, denoted by \( \mathbf{s}_N \in \mathbb{C}^N \), we first construct two prime-length sequences of length \(Q_1\) and \(Q_2\) respectively. Then, we append \(Q_2\)-length sequence to the \(Q_1\)-length sequence to generate an \(N\)-length sequence as follows
\begin{align}
    s_N[m] = 
    \begin{cases} 
        s_{Q_1}[m], & \text{if } m = 0,\cdots,Q_1-1,\\
        s_{Q_2}[m-Q_1], & \text{if } m = Q_1,\cdots,N-1.
    \end{cases}
    \label{eq:cazac_even_math_rep}
\end{align}
\begin{figure*}[t]
% \begin{strip} % Adjust the space as needed
\begin{align}
    % \footnotesize
    % Your long here
        \mathbf{S}_N = \begin{bmatrix}
        \multicolumn{7}{c}{\text{Top block } \mathbf{S}_{Q_1}} \\[3pt]
        s_{Q_1}[0] & s_{Q_1}[Q_1-1] & \cdots & s_{Q_1}[Q_1-Q_2+1] & s_{Q_1}[Q_1-Q_2] & \cdots & s_{Q_1}[1]\\
        s_{Q_1}[1] & s_{Q_1}[0] & \cdots & s_{Q_1}[Q_1-Q_2+2] & s_{Q_1}[Q_1-Q_2+1] & \cdots & s_{Q_1}[2]\\
        \vdots & \vdots & \ddots & \vdots & \vdots & \ddots & \vdots\\
        s_{Q_1}[Q_1-1] & s_{Q_1}[Q_1-2] & \cdots & s_{Q_1}[Q_1-Q_2] & s_{Q_1}[Q_1-Q_2-1] & \cdots & s_{Q_1}[0]\\\\
        \hdashline\\[-4pt]
        \multicolumn{7}{c}{\text{Bottom block } \mathcal{I}(\mathbf{S}_{Q_2})} \\[3pt]
        s_{Q_2}[0] & s_{Q_2}[{Q_2}-1] & \cdots & s_{Q_2}[1] & s_{Q_2}[0] & \cdots & s_{Q_2}[n
        \mod Q_2]\\
        s_{Q_2}[1] & s_{Q_2}[0] & \cdots & s_{Q_2}[2] & s_{Q_2}[1] & \cdots & s_{Q_2}[(n+1)
        \mod Q_2]\\
        \vdots & \vdots & \ddots & \vdots & \vdots & \ddots & \vdots\\
        s_{Q_2}[{Q_2}-1] & s_{Q_2}[{Q_2}-2] & \cdots & s_{Q_2}[0] & s_{Q_2}[{Q_2}-1] & \cdots & s_{Q_2}[(n+Q_2-1) \mod Q_2]
    \end{bmatrix}
    \label{eq:non_prime_cazac_even_cs}
\end{align}
% \end{strip}
\vspace{-22pt}
\end{figure*}

Following the same pattern, we can generate multiple \(N\)-length sequences by appending the cyclically shifted versions of \(Q_2\)-length sequences to the cyclically shifted versions of \(Q_1\)-length sequences. Let \(\mathbf{S}_{Q_1} \in \mathbb{C}^{Q_1 \times Q_1},\,\mathbf{S}_{Q_2} \in \mathbb{C}^{Q_2 \times Q_2}\) denote circulant matrices whose columns correspond to cyclically shifted versions of base sequences of lengths \(Q_1\) and \(Q_2\), respectively. We can generate \(N\)-length sequences, denoted by \( \mathbf{S}_N \), by repeatedly appending the sequences from \( \mathbf{S}_{Q_2} \) to those of \( \mathbf{S}_{Q_1} \) in a structured manner. Each of the \( Q_1 \)-length sequences is extended by appending one of the \( Q_2 \)-length sequences. Once all \( Q_2 \)-length sequences are used, we repeat the \( Q_2 \)-length sequences until all \( Q_1 \)-length sequences have been extended. A representative construction of \( \mathbf{S}_N \) is illustrated in~\eqref{eq:non_prime_cazac_even_cs}, where \(\mathcal{I}(\mathbf{S}_{Q_2})\) represents the repeated versions of \( \mathbf{S}_{Q_2} \), where \(n = Q_2 + 1 - (Q_1\mod Q_2)\). In addition to cyclic shifts, the proposed construction naturally extends to sequences generated using different root indices. Let \(\mathbf{S}^r_{Q_1} \in \mathbb{C}^{Q_1 \times Q_1-1},\,\mathbf{S}^r_{Q_2} \in \mathbb{C}^{Q_2 \times Q_2-1}\) denote matrices whose columns correspond to sequences generated using distinct root indices for lengths \(Q_1\) and \(Q_2\), respectively. We can generate \(N\)-length sequences by repeatedly appending the sequences from \( \mathbf{S}^r_{Q_2} \) to those of \( \mathbf{S}^r_{Q_1} \) in a structured manner. %Each of the \( Q_1\)-length sequences is extended by appending one of the \( Q_2 \)-length sequences. Once all \( Q_2\)-length sequences are used, we repeat the \( Q_2 \)-length sequences until all \( Q_1-1\)-length sequences have been extended. 
A representative construction of \( \mathbf{S}^r_N\) follows the same structure as~\eqref{eq:non_prime_cazac_even_cs}, except that the columns are associated with different root indices and appending pattern repeats until \(Q_1-1\) sequences are exhausted as opposed to \(Q_1\) in~\eqref{eq:non_prime_cazac_even_cs}.
% \begin{align}
%     % \footnotesize
%     % Your long here
%         \mathbf{S}^r_N = \begin{bmatrix}
%         \multicolumn{7}{c}{\text{Top block } \mathbf{S}^r_{Q_1}} \\[3pt]
%         s^1_{Q_1}(0) & s^2_{Q_1}(0) &\cdots & s^{Q_1-1}_{Q_1}(1)\\
%         \vdots & \vdots & \ddots & \vdots\\
%         s^1_{Q_1}(Q_1-1) & s^2_{Q_1}(Q_1-1) & \cdots & s^{Q_1-1}_{Q_1}(0)\\\\
%         \hdashline\\[-4pt]
%         \multicolumn{7}{c}{\text{Bottom block } \mathcal{I}(\mathbf{S}^2_{Q_2})} \\[3pt]
%         s_{Q_2}(0) & s_{Q_2}({Q_2}-1) & \cdots & s_{Q_2}(1) & s_{Q_2}(0) & \cdots & s_{Q_2}(n
%         \mod Q_2)\\
%         s_{Q_2}(1) & s_{Q_2}(0) & \cdots & s_{Q_2}(2) & s_{Q_2}(1) & \cdots & s_{Q_2}((n+1)
%         \mod Q_2)\\
%         \vdots & \vdots & \ddots & \vdots & \vdots & \ddots & \vdots\\
%         s_{Q_2}({Q_2}-1) & s_{Q_2}({Q_2}-2) & \cdots & s_{Q_2}(0) & s_{Q_2}({Q_2}-1) & \cdots & \mathbf{s}_{Q_2}((n+Q_2-1) \mod Q_2)
%     \end{bmatrix}
%     \label{eq:non_prime_cazac_even_dr}
% \end{align}

Notably, in the proposed construction, we limit the maximum number of sequences to \(\max\{Q_1,Q_2\}\) when extending cyclically shifted versions and \(\max\{Q_1-1,Q_2-1\}\) when extending different root index sequences. This is because extending beyond these limits leads to repetition of the underlying appending pattern, which in turn produces duplicate sequence sets. The construction of even-length sequences is particularly relevant for OFDM systems~\cite{3gpp::38211}, where \( N \) is typically an even number and a multiple of \(12\). It is important to note that the construction pattern presented in~\eqref{eq:non_prime_cazac_even_cs} is for illustrative purposes, and there is no need to follow this specific appending pattern. For instance, instead of utilizing all \(Q_1\) sequences, one may select a subset of size \(Q_2\) consisting of evenly spaced cyclic shifts of the \(Q_1\)-length base sequence. The selected sequences can then be extended by appending cyclically shifted versions of the \(Q_2\)-length sequences. Such a selection preserves zero-lag cross-correlation (inner-product) orthogonality within the chosen subset. This approach is particularly useful in applications such as PRACH generation in 5G NR~\cite{3gpp::38211}, where cyclic shifts are configured to accommodate maximum delay uncertainty. A related design problem is discussed in Section~\ref{sec:bjorck_high_Doppler}. 

Having said that, the process of repeatedly appending \(Q_2\)-length sequences introduces overlap regions, wherein multiple columns of \(\mathbf{S}_N\) or \(\mathbf{S}^r_N\) share identical appended segments from \(\mathbf{S}_{Q_2}\) or \(\mathbf{S}^r_{Q_2}\). In particular, these overlapping portions become significant when \(Q_2\) is close to \(Q_1\). As a result, the zero-lag cross-correlation (inner product) between cyclically shifted sequences becomes non-zero, and the inner products between sequences generated using different root indices deviate from the constant values typically observed in prime-length CAZAC constructions. We discuss more on design choices of selecting \((Q_1,\,Q_2)\) in the subsequent parts of this section. We next present expressions for the inner product of the extended cyclically shifted sequences and for the extended different root indices, which further justify the above mentioned limits on the maximum number of usable sequences.
\begin{lemma}[Inner Product for Extending Cyclically Shifted Versions: Even $N$]\label{lem:inner_product_even_cyclic}
Consider two sequences $\mathbf{s}_i,\,\mathbf{s}_j \in \mathbb{C}^N$ of length \(N = Q_1 + Q_2\), constructed as
\begin{align}
s_i[n] =
\begin{cases}
x_i[n], & 0 \le n \le Q_1 - 1, \\[4pt]
x_{\pi(i)}[n-Q_1], & Q_1 \le n \le N-1,
\end{cases}
\end{align}
\begin{align}
s_j[n] =
\begin{cases}
x_j[n], & 0 \le n \le Q_1 - 1, \\[4pt]
x_{\pi(j)}[n-Q_1], & Q_1 \le n \le N-1,
\end{cases}
\end{align}
where $\mathbf{x}_i,\,\mathbf{x}_j \in \mathbb{C}^{Q_1}$ are $Q_1$-length sequences corresponding to cyclic shift indices $i$ and $j$, respectively, and $\mathbf{y}_{\pi(i)},\,\mathbf{y}_{\pi(j)} \in \mathbb{C}^{Q_2}$ are $Q_2$-length sequences whose cyclic shift indices are determined by a mapping $\pi(\cdot)$. Then, the absolute value of the normalized inner product between $\mathbf{s}_i$ and $\mathbf{s}_j$ is given by
\begin{align}
    \rho_{i,j} = \frac{1}{N}\left|\langle \mathbf{s}_i, \mathbf{s}_j \rangle\right| =
    \begin{cases}
    \dfrac{Q_1}{N}, & i=j\,\,\text{and}\,\,\pi(i)\neq\pi(j), \\[6pt]
    \dfrac{Q_2}{N}, & i\neq j\,\,\text{and}\,\,\pi(i)=\pi(j), \\[6pt]
    0, & i\neq j\,\,\text{and}\,\,\pi(i)\neq\pi(j).
    \end{cases}
    \label{eq:inner_product_cyclic_even}
    \end{align}
\end{lemma}

\begin{IEEEproof}
By construction, the inner product between \(\mathbf{s}_i\) and \(\mathbf{s}_j\) decomposes as
\begin{align}
\left\langle \mathbf{s}_i, \mathbf{s}_j\right\rangle
&=
\sum_{n=0}^{N-1}s_i[n]s_j^*[n] \nonumber\\
&=
\sum_{n=0}^{Q_1-1}x_{i}[n]x_{j}^*[n]
+
\sum_{m=0}^{Q_2-1}x_{\pi(i)}[m]x_{\pi(j)}^*[m].
\label{eq:proof_even_cyclic_split}
\end{align}

Using simple arithmetic operations, we can obtain the expressions in~\eqref{eq:inner_product_cyclic_even} from~\eqref{eq:proof_even_cyclic_split}.
\end{IEEEproof}

The result in Lemma~\ref{lem:inner_product_even_cyclic} highlights an important design insight: it is desirable to construct sequences with no overlap in their appended segments. Specifically, length-\(N\) sequences that do not share either the \(Q_1\)-length or \(Q_2\)-length sequences derived from \(\mathbf{S}_{Q_1}\) and \(\mathbf{S}_{Q_2}\), respectively, achieve perfect orthogonality. However, enforcing such a non-overlapping constraint inherently limits the maximum number of mutually orthogonal sequences to \(\min\{Q_1,Q_2\}\). To illustrate this tradeoff, consider the example \(N=120\). Selecting the largest prime less than \(N\) yields \((Q_1,Q_2)=(113,7)\). Using the construction in~\eqref{eq:non_prime_cazac_even_cs}, up to \(Q_1=113\) sequences of length \(N\) can be generated through repeated appending of \(Q_2\)-length sequences. However, the largest set of mutually orthogonal sequences is limited to only \(7\) (with no overlapping patterns). If a limited overlap of \(7\) samples is acceptable, all \(113\) sequences may still be utilized, similar to the repetition-based construction in~\eqref{eq:non_prime_seq_rep}. Alternatively, the flexibility of the proposed framework allows selecting different prime pairs \((Q_1,\,Q_2)\) to increase the number of orthogonal sequences. For instance, choosing \((Q_1, Q_2) = (61, 59)\) enables up to \(59\) mutually orthogonal sequences by restricting the usable set to \(59\) sequences. However, this comes at the cost of reducing the total number of available sequences. This highlights a key advantage of the proposed construction: the ability to trade off between the total number of sequences and the number of orthogonal sequences.

%In particular, choosing a larger \(Q_1\) (e.g., the largest prime less than \(N\)) maximizes the total number of sequences but limits orthogonality, whereas more balanced choices of \((Q_1,Q_2)\) improve orthogonality at the expense of the total sequence count. 
Overall, Lemma~\ref{lem:inner_product_even_cyclic} provides clear justification for limiting the maximum number of generated sequences to \(\max\{Q_1,Q_2\}\), especially when \(\max\{Q_1,Q_2\}\gg \min\{Q_1,Q_2\}\). Generating additional sequences beyond this limit results in repeated appending patterns, which increases the normalized inner product beyond \({\max\{Q_1,Q_2\}}/{N}\) and can eventually approach \(1\), thereby defeating the objective of constructing sequences with minimal to no overlap (i.e., near or perfect orthogonal sequences). In summary, if perfect orthogonality is the primary objective, one should select \((Q_1, Q_2)\) such that both primes are close to each other and limit the maximum number of sequences to \(\min\{Q_1, Q_2\}\). Conversely, if maximizing the total number of sequences is more important, selecting a larger \(Q_1\) is beneficial, although at the cost of increased inner product among the resulting sequences.

We now present below the inner product for the extended versions of different root indices. For prime-length CAZAC families whose different-root inner products have constant magnitude (e.g., ZC sequences), the following result characterizes the inner product behavior for even-length sequences.

\begin{lemma}[Inner Product for Extended Different-Root Sequences: Even \(N\)]
\label{lem:inner_product_even_diff_roots}
Let \(N=Q_1+Q_2\), where \(Q_1\) and \(Q_2\) are prime numbers. Consider two length-\(N\) sequences \(\mathbf{s}_{(u,v)}\) and \(\mathbf{s}_{(u',v')}\) constructed as
\begin{align}
s_{(u,v)}[n]
=
\begin{cases}
x_{u}[n], & 0 \le n \le Q_1-1,\\[4pt]
x_{v}[n-Q_1], & Q_1 \le n \le N-1,
\end{cases}
\end{align}
and
\begin{align}
s_{(u',v')}[n]
=
\begin{cases}
x_{u'}[n], & 0 \le n \le Q_1-1,\\[4pt]
x_{v'}[n-Q_1], & Q_1 \le n \le N-1,
\end{cases}
\end{align}
where \(\mathbf{x}_{u},\,\mathbf{x}_{u'} \in \mathbb{C}^{Q_1} \) are \(Q_1\)-length CAZAC sequences generated using root indices \(u\) and \(u'\), respectively, and \(\mathbf{y}_{v},\,\mathbf{y}_{v'} \in \mathbb{C}^{Q_2}\) are \(Q_2\)-length CAZAC sequences generated using root indices \(v\) and \(v'\), respectively.

Assume that, for the underlying prime-length CAZAC family, the inner product between two distinct root-index sequences of length \(Q\) has constant magnitude \(\sqrt{Q}\) (e.g., ZC sequences~\cite{gregoratti2023mathematicalpropertieszadoffchusequences}), i.e.,
\begin{align}
\sum_{n=0}^{Q-1} x_{r}[n]x_{r'}^*[n]
=
\begin{cases}
Q, & r=r',\\
\sqrt{Q}\,e^{j\phi(r,r')}, & r\neq r',
\end{cases}
\label{eq:generic_cazac_root_assumption}
\end{align}
where \(\mathbf{x}_{r},\,\mathbf{x}_{r'}\in \mathbb{C}^{Q}\) are two CAZAC sequences with root indices \(r\) and \(r'\), respectively, and \(\phi(r,r')\) is some phase which is typically a function of \((r-r')\).

Define the magnitude of the normalized inner product
\begin{align}
\rho_{(u,v),(u',v')}
=
\frac{1}{N}\left|\left\langle \mathbf{s}_{(u,v)},\mathbf{s}_{(u',v')}\right\rangle\right|.
\label{eq:norm_inner_product_diff_roots}
\end{align}

Then, the following bounds hold for different cases.

1) \(u=u'\) and \(v\neq v'\):
\begin{align}
\frac{|Q_1-\sqrt{Q_2}|}{N}
\le
\rho_{(u,v),(u',v')}
\le
\frac{Q_1+\sqrt{Q_2}}{N}.
\label{eq:inner_product_diff_roots_even_2}
\end{align}

2) \(u\neq u'\) and \(v=v'\):
\begin{align}
\frac{|Q_2-\sqrt{Q_1}|}{N}
\le
\rho_{(u,v),(u',v')}
\le
\frac{Q_2+\sqrt{Q_1}}{N}.
\label{eq:inner_product_diff_roots_even_3}
\end{align}

3) \(u\neq u'\) and \(v\neq v'\):
\begin{align}
\frac{|\sqrt{Q_1}-\sqrt{Q_2}|}{N}
\le
\rho_{(u,v),(u',v')}
\le
\frac{\sqrt{Q_1}+\sqrt{Q_2}}{N}.
\label{eq:inner_product_diff_roots_even_4}
\end{align}
\end{lemma}

\begin{IEEEproof}
By construction, the inner product between \(\mathbf{s}_{(u,v)}\) and \(\mathbf{s}_{(u',v')}\) decomposes as
\begin{align}
\left\langle \mathbf{s}_{(u,v)}, \mathbf{s}_{(u',v')}\right\rangle
&=
\sum_{n=0}^{N-1}s_{(u,v)}[n]s_{(u',v')}^*[n] \nonumber\\
&=
\sum_{n=0}^{Q_1-1}x_{u}[n]x_{u'}^*[n]
+
\sum_{m=0}^{Q_2-1}x_{v}[m]x_{v'}^*[m].
\label{eq:proof_even_diff_roots_split}
\end{align}

Using the assumption in~\eqref{eq:generic_cazac_root_assumption}, we have
\begin{align}
\sum_{n=0}^{Q_1-1}x_u[n]x_{u'}^*[n]
=
\begin{cases}
Q_1, & u=u',\\[3pt]
\sqrt{Q_1}\,
e^{j\phi_1(u,u')}, & u\neq u',
\end{cases}
\label{eq:proof_even_roots_q1}
\end{align}
for some phase \(\phi_1(u,u')\). Similarly,
\begin{align}
\sum_{m=0}^{Q_2-1}x_v[m]x_{v'}^*[m]
=
\begin{cases}
Q_2, & v=v',\\[3pt]
\sqrt{Q_2}\,
e^{j\phi_2(v,v')}, & v\neq v',
\end{cases}
\label{eq:proof_even_roots_q2}
\end{align}
for some phase \(\phi_2(v,v')\).

Substituting into~\eqref{eq:proof_even_diff_roots_split} yields
\begin{align}
\left\langle \mathbf{s}_{(u,v)},\mathbf{s}_{(u',v')}\right\rangle
&=
\delta_{u,u'}Q_1
+
(1-\delta_{u,u'})\sqrt{Q_1}e^{j\phi_1(u,u')}
\nonumber\\
&\quad+
\delta_{v,v'}Q_2
+
(1-\delta_{v,v'})\sqrt{Q_2}e^{j\phi_2(v,v')},
\label{eq:proof_even_roots_master}
\end{align}
where \(\delta_{a,b}\) is the Kronecker delta. Evaluating the magnitude values of~\eqref{eq:proof_even_roots_master} for all three different cases and substituting into~\eqref{eq:norm_inner_product_diff_roots}, we obtain~\eqref{eq:inner_product_diff_roots_even_2}--\eqref{eq:inner_product_diff_roots_even_4}.
\end{IEEEproof}

From Lemma~\ref{lem:inner_product_even_diff_roots}, Case~2 (\(u\neq u'\), \(v=v'\)) and Case~3 (\(u\neq u'\), \(v\neq v'\)) provide useful insight into the scaling behavior of the normalized inner product with respect to \(N\), as summarized below.
\begin{remark}[Asymptotic Inner-Product Behavior for Different Roots]
\label{rem:inner_product_scaling}

\textit{Case 2: \(u\neq u'\), \(v=v'\).} The normalized inner product satisfies
\begin{align}
\frac{\left|Q_2-\sqrt{Q_1}\right|}{N}
\le
\rho_{(u,v),(u',v')}
\le
\frac{Q_2+\sqrt{Q_1}}{N}.
\end{align}
When \(Q_1 \gg Q_2\) and \(Q_1 \approx N\), both bounds satisfy
\begin{align}
\frac{\left|Q_2-\sqrt{Q_1}\right|}{N}
\approx
\frac{Q_2+\sqrt{Q_1}}{N}
\approx
\frac{1}{\sqrt{N}},
\end{align}
which implies
\begin{align}
\rho_{(u,v),(u',v')} \approx \frac{1}{\sqrt{N}}.
\end{align}

This behavior is analogous to prime-length CAZAC sequences, where the normalized inner product between different root indices scales as \(1/\sqrt{Q}\). In contrast, when \(Q_1 \approx Q_2 \approx N/2\), both bounds approach \(1/2\) (for large \(N\)), which is undesirable from an orthogonality point of view. Hence, for Case~2, choosing \(Q_1 \gg Q_2\) is desirable to achieve lower inner-product values.

\textit{Case 3: \(u\neq u'\), \(v\neq v'\).} The normalized inner product satisfies
\begin{align}
\frac{\left|\sqrt{Q_1}-\sqrt{Q_2}\right|}{N}
\le
\rho_{(u,v),(u',v')}
\le
\frac{\sqrt{Q_1}+\sqrt{Q_2}}{N}.
\end{align}
When \(Q_1 \gg Q_2\) and \(Q_1 \approx N\), both bounds satisfy
\begin{align}
\frac{\left|\sqrt{Q_1}-\sqrt{Q_2}\right|}{N}
\approx
\frac{\sqrt{Q_1}+\sqrt{Q_2}}{N}
\approx
\frac{1}{\sqrt{N}},
\end{align}
which again implies
\begin{align}
\rho_{(u,v),(u',v')} \approx \frac{1}{\sqrt{N}}.
\end{align}

When \(Q_1 \approx Q_2 \approx N/2\), we obtain
\begin{align}
\frac{\left|\sqrt{Q_1}-\sqrt{Q_2}\right|}{N} \approx 0,
\qquad
\frac{\sqrt{Q_1}+\sqrt{Q_2}}{N} \approx \sqrt{\frac{2}{N}},
\end{align}
which implies
\begin{align}
0 \;\lesssim\; \rho_{(u,v),(u',v')} \;\lesssim\; \sqrt{\frac{2}{N}}.
\end{align}

In summary, irrespective of the specific choice of \((Q_1,\,Q_2)\), the normalized inner product between sequences generated using different root indices satisfies
\begin{align}
\rho_{(u,v),(u',v')}
= \mathcal{O}\!\left(\frac{1}{\sqrt{N}}\right),
\end{align}
demonstrating that the proposed construction preserves the favorable scaling behavior of prime-length CAZAC sequences, while allowing flexibility in selecting \((Q_1,\,Q_2)\) to control the tightness of the bounds.
\end{remark}

The results in Lemma~\ref{lem:inner_product_even_diff_roots} provide insights analogous to those in Lemma~\ref{lem:inner_product_even_cyclic}, indicating that minimum or no overlap in the appended segments is beneficial in minimizing zero-lag cross-correlation (inner product). However, unlike the cyclic-shift-based construction, where the choice of \((Q_1,\,Q_2)\) determines the number of mutually orthogonal sequences through a strict non-overlapping constraint, the extension based on different root indices exhibits inherently bounded inner products even without enforcing strict orthogonality. In particular, as shown in Remark~\ref{rem:inner_product_scaling}, the normalized inner product between sequences generated using different root indices scales on the order of \(1/\sqrt{N}\), regardless of the specific choice of \((Q_1,\,Q_2)\). Thus, unlike the cyclic-shift case, the role of \((Q_1,\,Q_2)\) is not to enforce orthogonality, but rather to control the tightness of the inner-product bounds. More specifically, choosing \(Q_1 \gg Q_2\) ensures that the normalized inner product remains close to \(1/\sqrt{N}\), aligning with the behavior of prime-length CAZAC sequences. On the other hand, choosing \(Q_1 \approx Q_2\) and non-overlapping patterns helps extend both cyclically shifted versions and different root-index sequences to exhibit inner-product behavior.

{\em Generic construction for odd-length sequences.} For completeness, we extend the proposed construction to odd-length sequences. When \(N\) is odd, Goldbach's conjecture allows expressing \(N\) as \(N = Q_1 + Q_2 + Q_3\), where each \(Q_i\) is prime. For the construction based on cyclically shifted versions, we form circulant matrices 
\(\mathbf{S}_{Q_1}\in \mathbb{C}^{Q_1 \times Q_1}\), 
\(\mathbf{S}_{Q_2}\in \mathbb{C}^{Q_2 \times Q_2}\), and 
\(\mathbf{S}_{Q_3}\in \mathbb{C}^{Q_3 \times Q_3}\), 
whose columns correspond to cyclic shifts of base sequences of lengths \(Q_1\), \(Q_2\), and \(Q_3\), respectively. Similarly, for extending sequences corresponding to different root indices, we construct 
\(\mathbf{S}_{Q_1}^r\in \mathbb{C}^{Q_1 \times (Q_1-1)}\), 
\(\mathbf{S}_{Q_2}^r\in \mathbb{C}^{Q_2 \times (Q_2-1)}\), and 
\(\mathbf{S}_{Q_3}^r\in \mathbb{C}^{Q_3 \times (Q_3-1)}\). Without loss of generality, let \( Q_1 = \max\{Q_1, Q_2, Q_3\} \). Then, \(Q_1\) (for cyclic shifts) or \(Q_1-1\) (for different root indices) sequences of length \(N\) are constructed by repeatedly appending columns from \(\mathbf{S}_{Q_2}\) and \(\mathbf{S}_{Q_3}\), or \(\mathbf{S}_{Q_2}^r\) and \(\mathbf{S}_{Q_3}^r\), to the columns of \(\mathbf{S}_{Q_1}\) or \(\mathbf{S}_{Q_1}^r\), respectively. The resulting sequence matrix can be expressed as
\begin{align}
    \mathbf{S}_N =
    \begin{bmatrix}
        \mathbf{S}_{Q_1}\ \text{or}\ \mathbf{S}_{Q_1}^r \\[3pt]
        \hdashline \\[-8pt]
        \mathcal{I}\!\left(\mathbf{S}_{Q_2}\ \text{or}\ \mathbf{S}_{Q_2}^r\right) \\[3pt]
        \hdashline \\[-8pt]
        \mathcal{I}\!\left(\mathbf{S}_{Q_3}\ \text{or}\ \mathbf{S}_{Q_3}^r\right)
    \end{bmatrix},
    \label{eq:non_prime_cazac_odd}
\end{align}
where \(\mathcal{I}(\cdot)\) denotes a column-wise repetition (or selection) operator that maps the columns of \(\mathbf{S}_{Q_2}\) or \(\mathbf{S}_{Q_2}^r\), and \(\mathbf{S}_{Q_3}\) or \(\mathbf{S}_{Q_3}^r\), to match the number of columns in the top block. Due to space constraints, we omit a detailed analysis of the correlation properties for odd-length sequences. Nevertheless, the key insights derived for even-length constructions--particularly the dependence of inner-product behavior on overlap across appended segments--extend to the odd-length case as well.

\begin{figure*}[t]
     \centering
     \resizebox{0.97\textwidth}{!}{% Resize to match text width
     \begin{subfigure}[b]{0.485\textwidth}
         \centering
        \includegraphics[width=\textwidth]{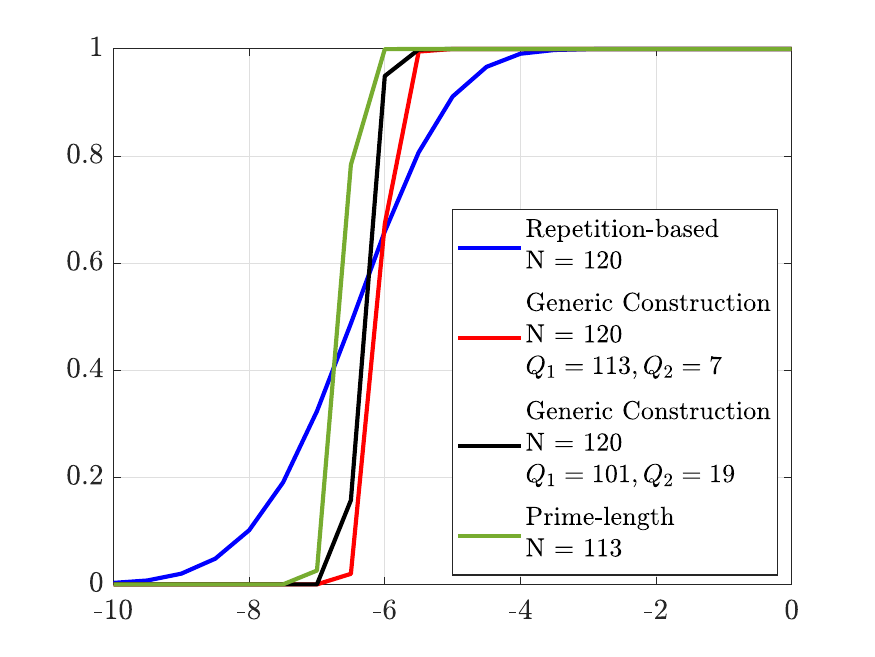}
        \captionsetup{font=footnotesize}
        \caption{{Prob. of Detection vs SINR.}}
        \label{fig:prob_of_detect_vs_sinr_tn}
     \end{subfigure}
     \hfill
      \begin{subfigure}[b]{0.485\textwidth}
         \centering
        \includegraphics[width=\textwidth]{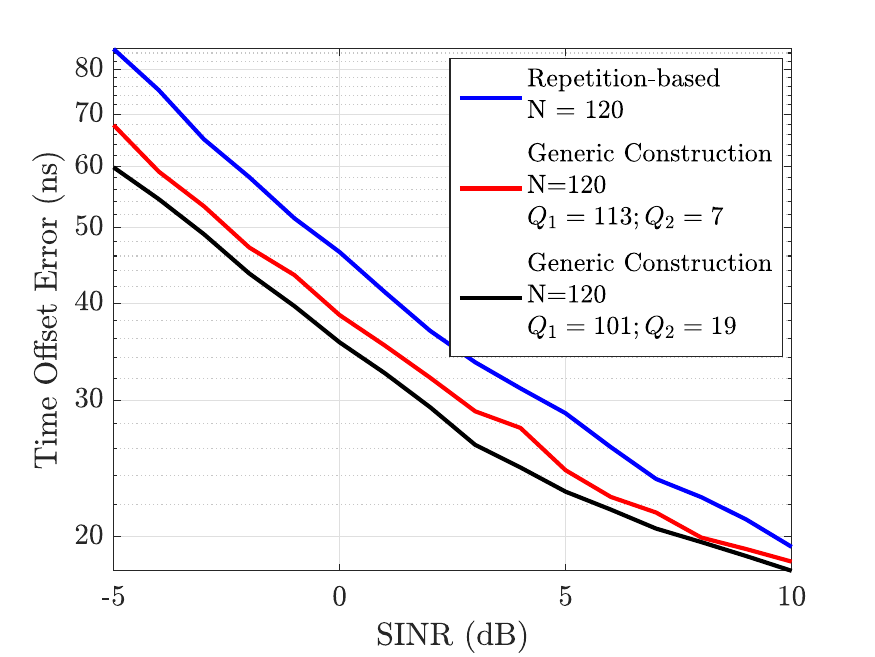}
        \captionsetup{font=footnotesize}
        \caption{Time Offset Error vs SINR.}
        \label{fig:time_ofst_err_vs_sinr_tn}
     \end{subfigure}
     }
     \captionsetup{font=small}
     \caption{{\em Probability of detection and time offset estimation error over different SINR values:} Performance comparison between inherent prime-length sequences, and extended sequences using the proposed construction and conventional repetition considering Bj\"orck sequences.}
     \label{fig:prob_of_detect_and_time_ofst_err_vs_sinr_tn}
     \vspace{-15pt}
\end{figure*}

To further demonstrate the advantages and flexibility of the proposed construction, we evaluate its performance in probability of detection and time-offset estimation and compare it with the conventional repetition-based extension of prime-length sequences. The objective of this study is to highlight the improved interference resilience offered by the proposed construction. For this evaluation, we consider delay and Doppler differences among interfering signals aligning with TN systems. For instance, in OFDM-based TN systems, propagation delay differences are contained within the cyclic prefix (CP), and Doppler differences are negligible relative to the SCS. \chb{As a result, the signals arrive within the CP in the time domain and within the SCS in the frequency domain, so they overlap completely at the receiver. Moreover, relative timing offsets between transmitters manifest as cyclic (modulo-\(N\)) shifts in the discrete-time domain.} We consider Bj\"orck sequences of length \(N=120\). The generation and correlation properties of the extended sequences are discussed in detail in the next section. Using Goldbach's conjecture for even integers, we consider two prime decompositions satisfying \(Q_1 + Q_2 = 120\): \((Q_1,\,Q_2)=(113,7)\) and \((Q_1,\,Q_2)=(101,19)\). 

Fig.~\ref{fig:prob_of_detect_vs_sinr_tn} compares the probability of detection for three cases: (i) prime-length Bj\"orck sequences, and non-prime length sequences constructed from Bj\"orck sequences (ii) using the proposed extension and (iii) via the conventional repetition-based approach, across different signal-to-interference-plus-noise ratio (SINR) values. The detection threshold is selected to achieve a false alarm probability of \(10^{-3}\) at an SINR of \(-5\) dB. Importantly, in this setup, the signal and noise powers are kept fixed, and only the interference power is varied across SINR values, thereby isolating the impact of interference. The results show that the proposed construction achieves a significantly higher probability of detection compared to the repetition-based approach, particularly in the low-SINR regime (corresponding to higher interference levels), with the prime-length case serving as a baseline. This improvement is due to the proposed construction preserving orthogonality among a subset of sequences even after extension, allowing interfering signals to be assigned to orthogonal resources and thereby reducing interference coupling as the signals overlap completely. Furthermore, the flexibility of the proposed construction in selecting different prime decompositions plays a critical role in performance. In particular, the configuration \((Q_1, Q_2) = (101, 19)\) achieves a higher probability of detection compared to \((113, 7)\), as it provides a larger set of mutually orthogonal sequences (\(19\) vs. \(7\)). This enables better interference mitigation and allows the performance to approach that of the prime-length baseline.

Extending the probability of detection analysis, Fig.~\ref{fig:time_ofst_err_vs_sinr_tn} compares the time-offset estimation error of the proposed construction with the repetition-based approach. In this case, we focus only on non-prime-length sequences, as they are more relevant for practical systems such as OFDM. Consistent with the probability-of-detection results, the proposed construction achieves lower estimation error across SINR values, with \((Q_1, Q_2) = (101, 19)\) outperforming \((113, 7)\), particularly under high interference conditions. These results highlight an important design insight: increasing the number of orthogonal sequences improves interference averaging and reduces worst-case interference coupling, leading to better performance across varying SINR conditions. In contrast, configurations with fewer orthogonal sequences exhibit greater overlap among interfering signals, leading to higher estimation errors, especially in the low-SINR regime.

\section{Bj\"orck Sequences and Their Applicability to TN and NTN Systems}
\subsection{Generation of Bj\"orck Sequences}
Bj\"orck sequences in their original form defined for prime lengths~\cite{Bjorck1990,BjrckAmbiguity4775877}. Given a prime number \(Q\), Bj\"orck sequences are expressed as
\begin{align}
    b_Q[m] = e^{j\theta_Q[m]}, \quad m = 0, 1, \cdots, Q-1,
\end{align}
where \(\theta_Q[m]\) is defined for two cases. For \(Q \equiv 1 \pmod{4}\), we have
\begin{align}
    \theta_Q[m] = \genfrac{(}{)}{}{}{m}{Q} \arccos\left(\frac{1}{1+\sqrt{Q}}\right),
    \label{eq:bjrck_case1}
\end{align}
and for \(Q \equiv 3 \pmod{4}\), we have
\begin{align}
    \theta_Q[m] = 
    \begin{cases} 
        \arccos\left(\frac{1-Q}{1+Q}\right), & \text{if } \genfrac{(}{)}{}{}{m}{Q} = -1, \\
        0, & \text{otherwise}.
    \end{cases}
    \label{eq:bjrck_case2}
\end{align}

Here, \(\genfrac{(}{)}{}{}{m}{Q}\) is a Legendre symbol defined as
\begin{align}
    \genfrac{(}{)}{}{}{m}{Q} =
    \begin{cases} 
        0, & \text{if } m \equiv 0 \pmod{Q}, \\
        1, & \text{if } m \text{ is a quadratic residue modulo }Q, \\
        -1, & \text{if } m \text{ is a quadratic nonresidue modulo }Q.
    \end{cases}
\end{align}

Bj\"orck sequences, unlike ZC sequences, do not support multiple sequences through different root indices. Instead, distinct sequences are generated via cyclic shifts of a single base sequence, and these cyclically shifted versions remain mutually orthogonal (in the periodic sense), owing to the CAZAC property. Accordingly, a set of \(Q\) orthogonal Bj\"orck sequences of length \(Q\) can be represented using a circulant matrix. Denoting the base sequence by \(\mathbf{b}_Q\) and the resulting circulant matrix by \(\mathbf{B}_Q\), we obtain
\begin{align}
    \mathbf{B}_Q = \begin{bmatrix}
        b_Q[0] & b_Q[Q-1] & \cdots & b_Q[1]\\
        b_Q[1] & b_Q[0] & \cdots & b_Q[2]\\
        \vdots & \vdots & \ddots & \vdots\\
        b_Q[Q-1] & b_Q[Q-2] & \cdots & b_Q[0]
    \end{bmatrix},
    \label{eq:circulant_bjorck_B^P_C}
\end{align}
where each column corresponds to a distinct cyclic shift of \(\mathbf{b}_Q\), forming a complete orthogonal set of Bj\"orck sequences. Note that Bj\"orck sequences satisfy a form of Fourier duality~\cite{fourierdualsbjorck}, wherein their Discrete Fourier Transform (DFT) also yields a CAZAC sequence (up to trivial transformations such as phase rotations and index permutations). This property, while shared by certain CAZAC constructions, is not universal across all CAZAC families. 

\subsection{Correlation Properties of Extended Prime-Length Bj\"orck Sequences}
Since Bj\"orck sequences are inherently defined for prime lengths, we employ the generic construction framework presented in Section~\ref{sec:generic_construction_cazac} to extend them to non-prime lengths. As noted earlier, Bj\"orck sequences do not support multiple sequences through different root indices; hence, distinct sequences are generated via cyclic shifts of a base sequence. Accordingly, we utilize the proposed construction to extend cyclically shifted versions, as described in~\eqref{eq:non_prime_cazac_even_cs}. It is important to note that, similar to the inner-product behavior analyzed in Section~\ref{sec:generic_construction_cazac}, extending cyclically shifted prime-length Bj\"orck sequences to non-prime lengths alters the underlying correlation properties. Specifically, while prime-length CAZAC sequences exhibit perfect periodic orthogonality, i.e., zero periodic cross-correlation at all nonzero lags for distinct cyclic shifts, this property is no longer strictly preserved after extension due to partial overlap between appended segments. To systematically quantify these effects, we analyze both periodic and aperiodic cross-correlation properties of the extended sequences via the proposed construction, using the root-mean-square (RMS) of the correlation as a metric.
\begin{lemma}[Periodic Cross-Correlation RMS for Extended Bj\"orck Sequences]
\label{lem:per_rms_extended_cyclic}
Consider two length-\(N\) sequences \(\mathbf{s}_i\) and \(\mathbf{s}_j\), with \(N=Q_1+Q_2\), constructed according to \eqref{eq:non_prime_cazac_even_cs} by appending \(Q_2\)-length cyclically shifted sequences to \(Q_1\)-length cyclically shifted sequences, i.e.,
\begin{align}
s_i[n]
=
\begin{cases}
x_i[n], & 0 \le n \le Q_1-1,\\[3pt]
y_{\pi(i)}[n-Q_1], & Q_1 \le n \le N-1,
\end{cases}
\end{align}
and
\begin{align}
s_j[n]
=
\begin{cases}
x_j[n], & 0 \le n \le Q_1-1,\\[3pt]
y_{\pi(j)}[n-Q_1], & Q_1 \le n \le N-1,
\end{cases}
\end{align}
where \(x_i,x_j\) are cyclically shifted versions of a base length-\(Q_1\) prime-length Bj\"rock sequence, and \(y_{\pi(i)},y_{\pi(j)}\) are cyclically shifted versions of a base length-\(Q_2\) prime-length Bj\"rock sequence selected by the mapping \(\pi(\cdot)\).

Let
\begin{align}
C_{ij}(\tau)
&=
\frac{1}{N}\sum_{n=0}^{N-1} s_i[n]\;s_j^*[(n-\tau)\bmod N],\,\tau=0,1,\dots,N-1,
\label{eq:lemma_per_def_Cij}
\end{align}
denote the normalized periodic cross-correlation. 

The RMS of the normalized periodic cross-correlation over one period is defined as
\begin{align}
\mathrm{RMS}_{ij}
=
\sqrt{\frac{1}{N}\sum_{\tau=0}^{N-1} |C_{ij}(\tau)|^2 }.
\label{eq:lemma_per_rms_def}
\end{align}

To accommodate favorable inner-product properties, we consider only cases with \(i\neq j\), i.e., we reuse only shorter sequences for repeated appending. Then, the following holds.

\textit{Case 1: \(\pi(i)\neq \pi(j)\).} Under the random-phase approximation for nonzero lags,
\begin{align}
\mathrm{RMS}_{ij}
\approx
\frac{1}{\sqrt{N}}\sqrt{1-\frac{1}{N}}
\approx
\frac{1}{\sqrt{N}}.
\label{eq:lemma_per_rms_case1}
\end{align}

\textit{Case 2: \(\pi(i)=\pi(j)\).} Under the same approximation for nonzero lags,
\begin{align}
\mathrm{RMS}_{ij}
\approx
\frac{1}{\sqrt{N}}
\sqrt{1-\frac{1}{N}+\left(\frac{Q_2}{N}\right)^2 }.
\label{eq:lemma_per_rms_case2}
\end{align}

Equivalently, relative to the benchmark \(1/\sqrt{N}\), the RMS scales as
\begin{align}
\frac{\mathrm{RMS}_{ij}}{1/\sqrt{N}}
\approx
\begin{cases}
\sqrt{1-\dfrac{1}{N}}, & \pi(i)\neq \pi(j),\\[8pt]
\sqrt{1-\dfrac{1}{N}+\left(\dfrac{Q_2}{N}\right)^2}, & \pi(i)=\pi(j).
\end{cases}
\label{eq:lemma_per_rms_ratio}
\end{align}

Thus, when \(Q_2\ll N\), both cases reduce approximately to \(1/\sqrt{N}\), whereas when \(Q_2\) is a non-negligible fraction of \(N\), the second case exhibits a noticeable RMS inflation due to the deterministic overlap at \(\tau=0\).
\end{lemma}

\begin{IEEEproof}
See Appendix~\ref{app::periodic_cross_corr}.
\end{IEEEproof}

From Lemma~\ref{lem:per_rms_extended_cyclic}, the two asymptotic regimes \(Q_1\gg Q_2\) and \(Q_1\approx Q_2\approx N/2\) provide useful insight into the behavior of the periodic cross-correlation RMS, as summarized below.

\begin{remark}[Asymptotic RMS Behavior for Extended Bj\"orck Sequences]
\label{rem:per_rms_scaling_cyclic}
Consider the RMS expressions in Lemma~\ref{lem:per_rms_extended_cyclic} for two distinct extended cyclically shifted sequences.

\textit{Case 1: \(\pi(i)\neq \pi(j)\).}
In this case,
\begin{align}
\mathrm{RMS}_{ij}
\approx
\frac{1}{\sqrt{N}}\sqrt{1-\frac{1}{N}},
\end{align}
which is independent of the specific choice of \((Q_1,Q_2)\). Hence, for large \(N\),
\begin{align}
\mathrm{RMS}_{ij}\approx \frac{1}{\sqrt{N}}.
\end{align}

Thus, when the appended portions do not overlap, the proposed construction preserves the familiar \(1/\sqrt{N}\) RMS scaling.

\textit{Case 2: \(\pi(i)=\pi(j)\).}
In this case,
\begin{align}
\mathrm{RMS}_{ij}
\approx
\frac{1}{\sqrt{N}}
\sqrt{1-\frac{1}{N}+\left(\frac{Q_2}{N}\right)^2}.
\end{align}

The effect of the appended overlap is reflected by the ratio \(Q_2/N\).

When \(Q_1\gg Q_2\) and \(Q_1\approx N\), we have \(Q_2/N\ll 1\), and hence
\begin{align}
\mathrm{RMS}_{ij}
\approx
\frac{1}{\sqrt{N}}
\sqrt{1-\frac{1}{N}}
\approx
\frac{1}{\sqrt{N}}.
\end{align}

Therefore, in the case of \(Q_1\gg Q_2\), even when two sequences share the same appended portion, the RMS remains close to the benchmark \(1/\sqrt{N}\).

On the other hand, when \(Q_1\approx Q_2\approx N/2\), we obtain
\begin{align}
\frac{Q_2}{N}\approx \frac{1}{2},
\end{align}
and therefore
\begin{align}
\mathrm{RMS}_{ij}
\approx
\frac{1}{\sqrt{N}}
\sqrt{1-\frac{1}{N}+\frac{1}{4}}
\approx
\frac{\sqrt{5}}{2\sqrt{N}},
\end{align}
for large \(N\). Thus, compared to the non-overlapping case, the RMS exhibits a noticeable inflation due to the deterministic \(\tau=0\) contribution.

In summary, the regime \(Q_1\gg Q_2\) minimizes the RMS inflation caused by overlap in the appended segment, whereas the regime \(Q_1\approx Q_2\) increases the number of available orthogonal sequences but also increases the RMS penalty when overlaps occur.
\end{remark}

\begin{lemma}[Aperiodic Cross-Correlation RMS for Extended Bj\"orck Sequences]
\label{lem:aper_rms_extended_cyclic}
Define the normalized aperiodic cross-correlation as
\begin{align}
C_{ij}^{(a)}(\tau)=
\frac{1}{N}R_{ij}^{(a)}(\tau)
\label{eq:aper_lemma_def}
\end{align}
where
\begin{align}
R_{ij}^{(a)}(\tau)=
\sum_{n=0}^{N-1-\tau} s_i[n+\tau]s_j^*[n],
\quad 0\le \tau \le N-1,
\end{align}
with the negative-lag symmetry
\begin{align}
R_{ij}^{(a)}(-\tau)=\big(R_{ji}^{(a)}(\tau)\big)^*.
\qquad \tau>0.
\label{eq:aper_lemma_sym}
\end{align}

Further, define the RMS of the normalized aperiodic cross-correlation over all lags as
\begin{align}
\mathrm{RMS}_{ij}^{(a)}
=
\sqrt{
\frac{1}{2N-1}
\sum_{\tau=-(N-1)}^{N-1}
\left| C_{ij}^{(a)}(\tau) \right|^2 }.
\label{eq:aper_lemma_rms}
\end{align}

We consider cases corresponding to \(i\neq j\). Then, the following holds.

\textit{Case 1: \(\pi(i)\neq \pi(j)\).} Under the random-phase approximation for nonzero lags,
\begin{align}
\mathbb{E}\!\left[\mathrm{MS}_{ij}^{(a)}\right]
\approx
\frac{
2\big(Q_2N+Q_1(Q_1-Q_2-1)\big)
}{
(2N-1)N^2
},
\label{eq:aper_case1_ms}
\end{align}
where \(\mathrm{MS}_{ij}^{(a)}= (\mathrm{RMS}_{ij}^{(a)})^2\).

\textit{Case 2: \(\pi(i)=\pi(j)\).} Under the same approximation,
\begin{align}
\mathbb{E}\!\left[\mathrm{MS}_{ij}^{(a)}\right]
\approx
\frac{
Q_2^2+2\big(Q_2N+Q_1(Q_1-Q_2-1)\big)
}{
(2N-1)N^2
}.
\label{eq:aper_case2_ms}
\end{align}

Let \(P = Q_2N+Q_1(Q_1-Q_2-1)\), then for large \(N\),
\begin{align}
\mathbb{E}\!\left[\mathrm{MS}_{ij}^{(a)}\right]
&\approx \frac{2P}{(2N-1)N^2}
\approx \frac{P}{N^3},
\qquad \pi(i)\neq \pi(j),
\label{eq:aper_case1_ms_simple}
\\
\mathbb{E}\!\left[\mathrm{MS}_{ij}^{(a)}\right]
&\approx \frac{Q_2^2+2P}{(2N-1)N^2},
\qquad \pi(i)=\pi(j),
\label{eq:aper_case2_ms_simple}
\end{align}
and since the dominant contribution in \(P\) scales as \(Q_1^2\), the RMS scales as
\begin{align}
\mathbb{E}\!\left[\mathrm{RMS}_{ij}^{(a)}\right]
\approx
\frac{Q_1}{N}\cdot \frac{1}{\sqrt{N}}
=
\left(1-\frac{Q_2}{N}\right)\frac{1}{\sqrt{N}}
<
\frac{1}{\sqrt{N}}.
\label{eq:aper_rms_scaling_final}
\end{align}
\end{lemma}

\begin{IEEEproof}
See Appendix~\ref{app::aperiodic_cross_corr}.
\end{IEEEproof}

From Lemma~\ref{lem:aper_rms_extended_cyclic}, the two asymptotic regimes \(Q_1\gg Q_2\) and \(Q_1\approx Q_2\approx N/2\) provide useful insight into the behavior of the aperiodic cross-correlation RMS, as summarized below.

\begin{remark}[Asymptotic Aperiodic RMS Behavior for Extended Bj\"orck Sequences]
\label{rem:aper_rms_scaling_cyclic}
Consider the RMS expressions in Lemma~\ref{lem:aper_rms_extended_cyclic} for two distinct extended cyclically shifted sequences.

\textit{Case 1: \(\pi(i)\neq \pi(j)\).}
In this case,
\begin{align}
\mathbb{E}\!\left[\mathrm{MS}_{ij}^{(a)}\right]
\approx
\frac{
2\big(Q_2N+Q_1(Q_1-Q_2-1)\big)
}{
(2N-1)N^2
},
\end{align}
and hence
\begin{align}
\mathbb{E}\!\left[\mathrm{RMS}_{ij}^{(a)}\right]
\approx
\sqrt{
\frac{
2\big(Q_2N+Q_1(Q_1-Q_2-1)\big)
}{
(2N-1)N^2
}
}.
\label{eq:aper_rms_case1_rem}
\end{align}

When \(Q_1\gg Q_2\) and \(Q_1\approx N\), the dominant term satisfies
\begin{align}
Q_2N+Q_1(Q_1-Q_2-1)\approx N^2,
\end{align}
which gives
\begin{align}
\mathbb{E}\!\left[\mathrm{RMS}_{ij}^{(a)}\right]
\approx
\frac{1}{\sqrt{N}}.
\end{align}

On the other hand, when \(Q_1\approx Q_2\approx N/2\), we obtain
\begin{align}
Q_2N+Q_1(Q_1-Q_2-1)
\approx
\frac{N^2}{2},
\end{align}
and therefore
\begin{align}
\mathbb{E}\!\left[\mathrm{RMS}_{ij}^{(a)}\right]
\approx
\sqrt{\frac{1}{2N-1}}
\approx
\frac{1}{\sqrt{2N}}.
\end{align}

\textit{Case 2: \(\pi(i)=\pi(j)\).}
In this case,
\begin{align}
\mathbb{E}\!\left[\mathrm{MS}_{ij}^{(a)}\right]
\approx
\frac{
Q_2^2+2\big(Q_2N+Q_1(Q_1-Q_2-1)\big)
}{
(2N-1)N^2
},
\end{align}
and thus
\begin{align}
\mathbb{E}\!\left[\mathrm{RMS}_{ij}^{(a)}\right]
\approx
\sqrt{
\frac{
Q_2^2+2\big(Q_2N+Q_1(Q_1-Q_2-1)\big)
}{
(2N-1)N^2
}
}.
\label{eq:aper_rms_case2_rem}
\end{align}

When \(Q_1\gg Q_2\) and \(Q_1\approx N\), the zero-lag contribution \(Q_2^2\) is negligible compared to the dominant term \(Q_1^2\), and hence
\begin{align}
\mathbb{E}\!\left[\mathrm{RMS}_{ij}^{(a)}\right]
\approx
\frac{1}{\sqrt{N}}.
\end{align}

In contrast, when \(Q_1\approx Q_2\approx N/2\), we have
\begin{align}
Q_2^2+2\big(Q_2N+Q_1(Q_1-Q_2-1)\big)
\approx
\frac{5N^2}{4},
\end{align}
which yields
\begin{align}
\mathbb{E}\!\left[\mathrm{RMS}_{ij}^{(a)}\right]
\approx
\sqrt{\frac{5}{4(2N-1)}}
\approx
\frac{\sqrt{5}}{2\sqrt{2N}}.
\end{align}

Overall, unlike the periodic case, the aperiodic RMS can become smaller than the conventional \(1/\sqrt{N}\) benchmark because the effective overlap length decreases with lag. In particular, when \(Q_1\gg Q_2\), the RMS remains approximately \(1/\sqrt{N}\), whereas when \(Q_1\approx Q_2\approx N/2\), the RMS decreases to approximately \(1/\sqrt{2N}\) for \(\pi(i)\neq \pi(j)\) and to \(\sqrt{5}/(2\sqrt{2N})\) for \(\pi(i)=\pi(j)\). This shows that more balanced choices of \((Q_1,\,Q_2)\) can improve the average aperiodic cross-correlation behavior, even though they may introduce a larger deterministic zero-lag term when the appended portions overlap (not an issue if we choose sequences with no overlap).
\end{remark}

In summary, the correlation analysis reveals a fundamental distinction between the periodic and aperiodic regimes. For prime-length CAZAC sequences, distinct cyclic shifts are perfectly orthogonal under periodic correlation, yielding zero periodic cross-correlation. However, this ideal property is generally lost under non-prime-length extension due to the appended structure, which introduces overlap terms. Nevertheless, the RMS of the resulting periodic cross-correlation remains well controlled and scales on the order of \(1/\sqrt{N}\), with an additional zero-lag contribution only when the appended portions overlap. As noted earlier, periodic cross-correlation is relevant for TN scenarios, as signals from different transmitters fully overlap at the receiver due to negligible delay and Doppler differences. In contrast, aperiodic cross-correlation is nonzero even for prime-length cyclically shifted CAZAC sequences and typically scales on the order of \(1/\sqrt{N}\)~\cite{arasu2023bjorck}. The proposed construction preserves this favorable aperiodic scaling. Aperiodic cross-correlation is particularly important for NTN scenarios, where large delays and Doppler differences across satellites result in signals arriving partially overlapped at the UE, except for the cell-center UEs, where satellite transmission times are synchronized~\cite{dureppagari2024leo}. In such cases, aperiodic cross-correlation becomes the more relevant metric for interference and detection analysis.

\subsection{Time and Frequency Offset Estimation: Bj\"orck vs ZC}
In this section, we study time and frequency offset estimation of Bj\"orck sequences in comparison with that of ZC sequences, which are widely used for reference signals in the existing wireless systems, especially cellular systems. Additionally, we also explore the suitability of Bj\"orck sequences to high Doppler environments such as LEO-based NTN systems due to their superior ambiguity function characteristics compared to ZC sequences. Specifically, we assess joint time- and frequency-offset estimation in which a receiver attempts to detect reference signals (RSs) transmitted via a TN or NTN gNB to the receiver link. This involves the receiver locally generating the RS, referred to as \(\mathbf{x}_\text{rs}(\cdot)\), correlating it with the received signal, denoted by \(\mathbf{y}(\cdot)\), under different frequency hypotheses. Specifically, we define a set of frequency offsets \(\mathcal{D} = \{f_{1}, f_{2}, \ldots, f_{K}\}\), and peak detection is performed while accounting for each frequency offset. %Fig.~\ref{fig:joint_delay_Doppler} depicts the circuit for time-domain correlation at various frequency offsets for peak detection. 
Alternatively, the same procedure can be carried out in the frequency domain by taking the Fourier Transform at different delay values. Typically, \(\mathbf{y}(\cdot)\) is of the form
\begin{align}
    x(kT_s) = \mathbf{h}(kT_s) \ast x_\text{rs}(kT_s) + \mathbf{n}(kT_s),
    \label{eq:rx_sgl_model}
\end{align}
where \( k \) represents the index of the time domain sample, \(T_s\) denotes the sampling period, \( \mathbf{h}(\cdot) \) is the channel impulse response (CIR), \( \mathbf{x}_\text{rs} \) is the transmitted RS, \(\mathbf{h}(kT_s) \ast x_\text{rs}(kT_s)\) is the convolution between \( \mathbf{h}(\cdot) \) and \( \mathbf{x}_\text{rs}(\cdot) \) capturing the distortion caused by CIR, and \( \mathbf{n}(\cdot) \) is additive-white-Gaussian-noise (AWGN) added at the receiver. Note that \(T_s = 1/f_s\), where \(f_s\) is the sampling rate.

For this evaluation, we consider two representative scenarios corresponding to TNs and NTN systems. The first scenario models TN systems, where Doppler offsets are relatively small due to moderate relative velocities between the transmitter and receiver. Even under extreme mobility conditions such as high-speed train scenarios with speeds approaching 1000 km/h, the resulting Doppler shift at a carrier frequency of 2 GHz remains on the order of approximately 1 kHz. %Such Doppler ranges are typical for sub-6 GHz cellular deployments and are well within the operational assumptions of existing 5G NR synchronization procedures. 
The second scenario models LEO-based NTN systems, in which significantly higher Doppler shifts arise from the rapid orbital motion of satellites. For LEO satellites orbiting at an altitude of approximately 600 km, the observable Doppler shift at a carrier frequency of 2 GHz can reach 40--45 kHz. These large Doppler shifts pose substantial challenges for delay-Doppler estimation and synchronization, motivating the need for reference sequences with favorable ambiguity-function properties.
\begin{figure*}[t]
     \centering
     \resizebox{0.97\textwidth}{!}{% Resize to match text width
     \begin{subfigure}[b]{0.485\textwidth}
         \centering
        \includegraphics[width=\textwidth]{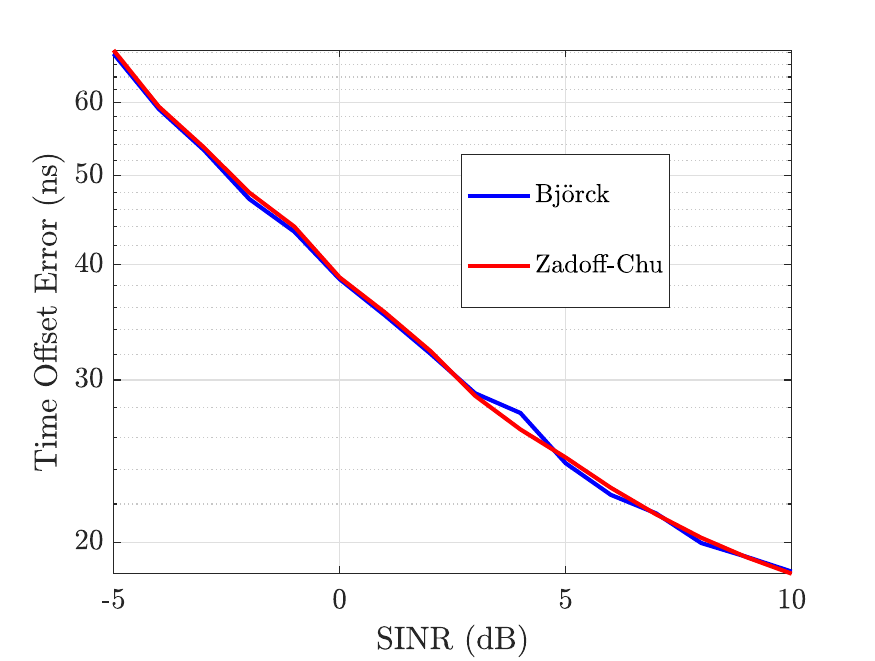}
        \captionsetup{font=footnotesize}
        \caption{{Time Offset Error vs SINR.}}
        \label{fig:time_ofst_err_vs_snr_tn}
     \end{subfigure}
     \hfill
      \begin{subfigure}[b]{0.485\textwidth}
         \centering
        \includegraphics[width=\textwidth]{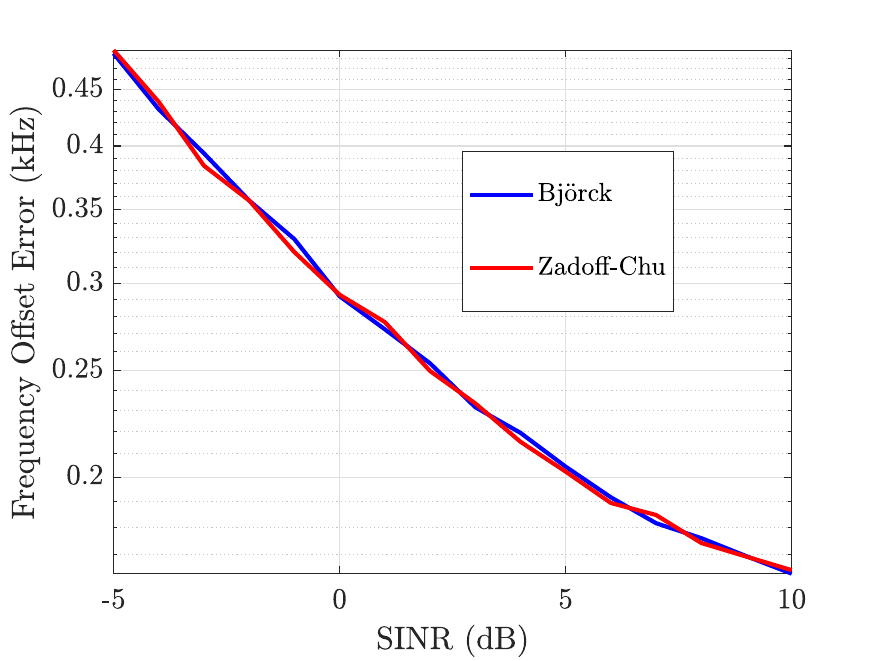}
        \captionsetup{font=footnotesize}
        \caption{Frequency Offset Error vs SINR.}
        \label{fig:freq_ofst_err_vs_snr_tn}
     \end{subfigure}
     }
     \captionsetup{font=small}
     \caption{{\em Performance comparison between ZC and Bj\"orck sequences for low Doppler environments such as TN systems:} {(a) Time offset and b) Frequency offset estimation error values over different SINRs.}}
     \label{fig:time_freq_ofst_err_vs_snr_tn}
     \vspace{-15pt}
\end{figure*}
\begin{figure*}[t]
     \centering
     \resizebox{0.97\textwidth}{!}{% Resize to match text width
     \begin{subfigure}[b]{0.485\textwidth}
         \centering
        \includegraphics[width=\textwidth]{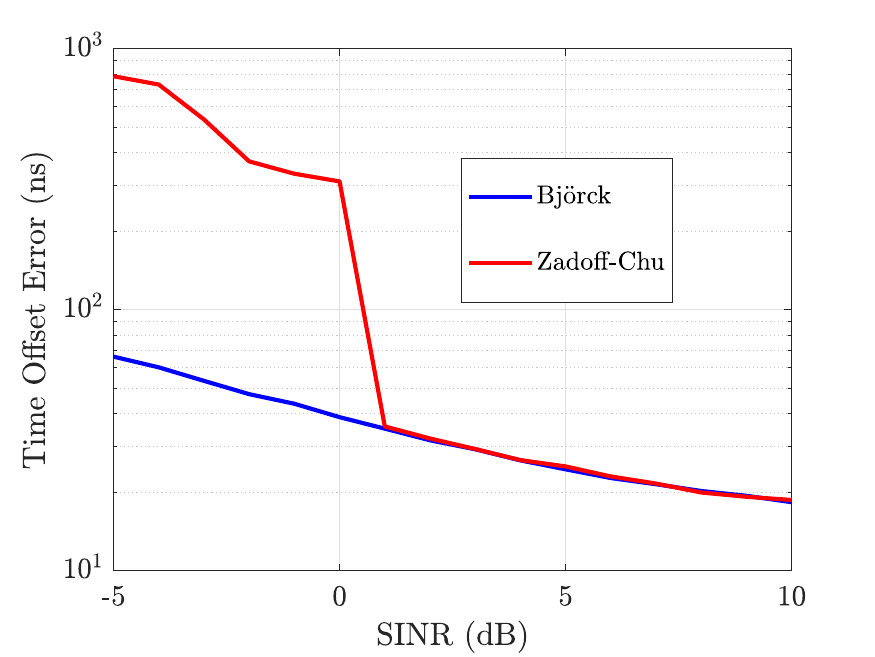}
        \captionsetup{font=footnotesize}
        \caption{{Time Offset Error vs SINR.}}
        \label{fig:time_ofst_err_vs_snr_ntn}
     \end{subfigure}
     \hfill
      \begin{subfigure}[b]{0.485\textwidth}
         \centering
        \includegraphics[width=\textwidth]{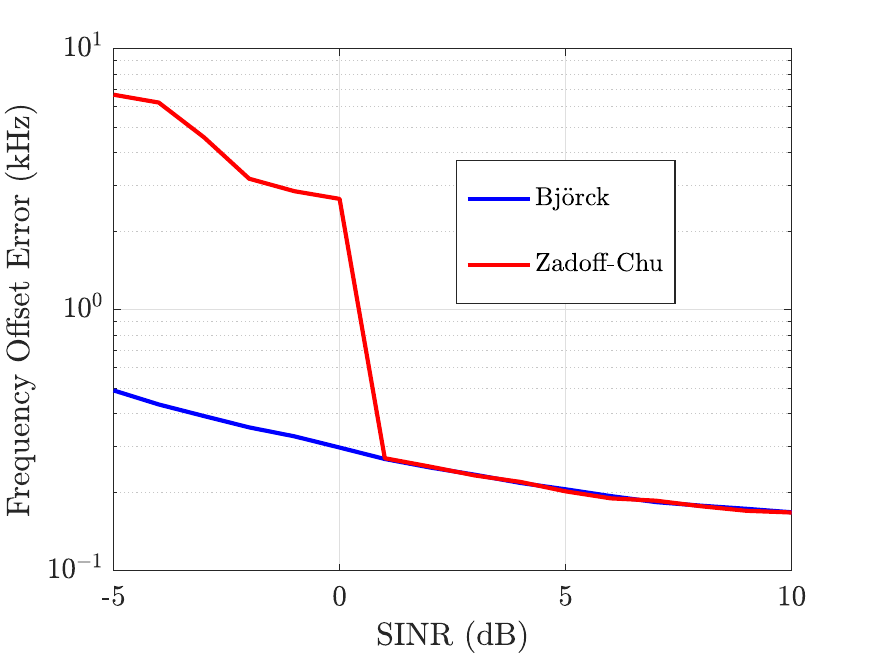}
        \captionsetup{font=footnotesize}
        \caption{Frequency Offset Error vs SINR.}
        \label{fig:freq_ofst_err_vs_snr_ntn}
     \end{subfigure}
     }
     \captionsetup{font=small}
     \caption{{\em Performance comparison between ZC and Bj\"orck sequences for high Doppler environments such as LEO-based NTN systems:} {(a) Time offset and b) Frequency offset estimation error values over different SINRs.}}
     \label{fig:time_freq_ofst_err_vs_snr_ntn}
     \vspace{-15pt}
\end{figure*}

To evaluate the performance of sequence designs under these conditions, we study both time-offset and frequency-offset estimation using Bj\"orck sequences and compare their performance with that of ZC sequences. For both sequence types, we consider a single OFDM symbol constructed from a sequence of length \(N=120\), corresponding to a \(2\) MHz bandwidth in OFDM-based systems with an SCS of 15 kHz. The receiver sampling rate is assumed to be 20 MHz. The carrier frequency is set to 2 GHz, which aligns with typical 5G NR sub-6 GHz deployments~\cite{3gpp::38901} and NTN evaluation studies~\cite{3gpp::38811,3gpp::38821}. For the low-Doppler TN scenario, frequency-offset estimation is performed using a set of frequency hypotheses spanning the range \([-2~\text{kHz},\, 2~\text{kHz}]\) with a step size of \(500\) Hz. For non-prime length extension, we use cyclically shifted versions for both Bj\"orck and ZC sequences. Fig.~\ref{fig:time_ofst_err_vs_snr_tn} and Fig.~\ref{fig:freq_ofst_err_vs_snr_tn} illustrate the resulting time-offset and frequency-offset estimation performance, respectively. As observed, in conventional terrestrial deployments, i.e., under low-Doppler conditions, Bj\"orck sequences achieve performance comparable to ZC sequences. This is consistent with the identical periodic cross-correlation behavior presented in Lemma~\ref{lem:per_rms_extended_cyclic}.
%Moreover, even when different root-index sequences are considered for ZC, a similar performance trend is expected, since the cross-correlation across root indices also scales on the order of \(1/\sqrt{N}\)~\cite{gregoratti2023mathematicalpropertieszadoffchusequences}, analogous to the periodic cross-correlation discussed in Lemma~\ref{lem:per_rms_extended_cyclic}. 
That said, ZC sequences remain advantageous in TN systems due to their support for multiple root-index sequences. It is also important to note that achievable time-offset estimation accuracy is primarily limited by the receiver sampling rate, whereas frequency-offset estimation accuracy is limited by the frequency-hypothesis step size. The system parameters used in this evaluation are chosen for demonstration purposes; although the absolute time and frequency performance numbers may vary with different parameters, the findings we present in this article remain unchanged.

Next, we evaluate the performance of time- and frequency-offset estimation under the high-Doppler conditions as in LEO-based NTN systems. In this scenario, Doppler shifts of up to 40 kHz are considered at a carrier frequency of 2 GHz. All other system parameters remain the same as in the TN scenario, except that the frequency-hypothesis search range is expanded to \([-45~\text{kHz},\,45~\text{kHz}]\) to account for the larger Doppler uncertainty. Fig.~\ref{fig:time_ofst_err_vs_snr_ntn} and Fig.~\ref{fig:freq_ofst_err_vs_snr_ntn} present the corresponding time-offset and frequency-offset estimation performance for the high-Doppler NTN scenario. While both Bj\"orck and ZC sequences exhibit comparable performance in low-Doppler environments, the advantages of Bj\"orck sequences become clearly evident under high-Doppler conditions. In particular, Bj\"orck sequences demonstrate improved estimation accuracy at low SINRs due to their superior ambiguity-function characteristics~\cite{BjrckAmbiguity4775877}, which yield sharper correlation peaks and reduced sidelobe levels in the presence of large Doppler shifts. At high SINRs, although ZC sequences produce additional false peaks, the desired peak remains distinguishable, resulting in performance comparable to that of Bj\"orck sequences. Furthermore, by comparing the results across both low-Doppler and high-Doppler scenarios, we observe that Bj\"orck sequences maintain consistent time- and frequency-offset estimation performance regardless of the Doppler range, highlighting their robustness to varying Doppler conditions. These results highlight the strong potential of Bj\"orck sequences as a viable alternative to ZC sequences for synchronization and reference signal design in future wireless systems. For instance, ZC sequences are currently employed for PRACH generation in 5G NR~\cite{3gpp::38211} and are also being considered as a baseline design for PRACH in emerging 6G studies. The improved Doppler robustness of Bj\"orck sequences suggests that they could serve as an effective replacement for ZC sequences in such applications, particularly in high Doppler environments, where reliable synchronization between gNBs and UEs is critical. In summary, we tabulate key strengths and weaknesses of Bj\"orck and ZC sequences under different aspects in Table~\ref{tab:seq_comparison} to facilitate sequence selection based on system requirements. 
\vspace{-7pt}
\begin{table}[htbp]
\centering
\caption{Comparison of Bj\"orck and ZC Sequences}
\label{tab:seq_comparison}
\renewcommand{\arraystretch}{1.1}
\begin{tabular}{|p{2.25cm}|p{2.5cm}|p{2.7cm}|}
\hline
\textbf{Aspect} & \textbf{Bj\"orck} & \textbf{ZC} \\ \hline
Ambiguity function & Reduced sidelobes; better peak detection & Distorted under high Doppler \\ \hline
Doppler robustness & High in NTN; cleaner ambiguity characteristics & Degrades under high Doppler (spurious peaks) \\ \hline
Low-Doppler performance & Comparable to ZC & Comparable to Bj\"orck \\ \hline
Time/frequency offset estimation & Consistent across different Doppler conditions & Degrades under high Doppler \\ \hline
Multiple sequences & Cyclic shifts only & Cyclic shifts and multiple root indices \\ \hline
Standardization & Limited exploration & Widely used in LTE/5G NR; baseline in 6G \\ \hline
\end{tabular}
\vspace{-7pt}
\end{table}

To facilitate the reproducibility of the evaluation framework and the corresponding numerical results, all simulation parameters and assumptions are aligned with 3GPP TN and NTN evaluation methodologies~\cite{3gpp::38901,3gpp::38811,3gpp::38821}. Furthermore, the sequence generation procedures, signal model, and evaluation metrics are explicitly specified to enable independent reproduction.
\section{Doppler-Depedent Behavior of Bj\"orck Sequences: NTN Relevance}\label{sec:bjorck_high_Doppler}
\begin{figure*}[t]
     \centering
     \resizebox{0.90\textwidth}{!}{% Resize to match text width
     \begin{subfigure}[b]{0.325\textwidth}
         \centering
         \includegraphics[width=\textwidth]{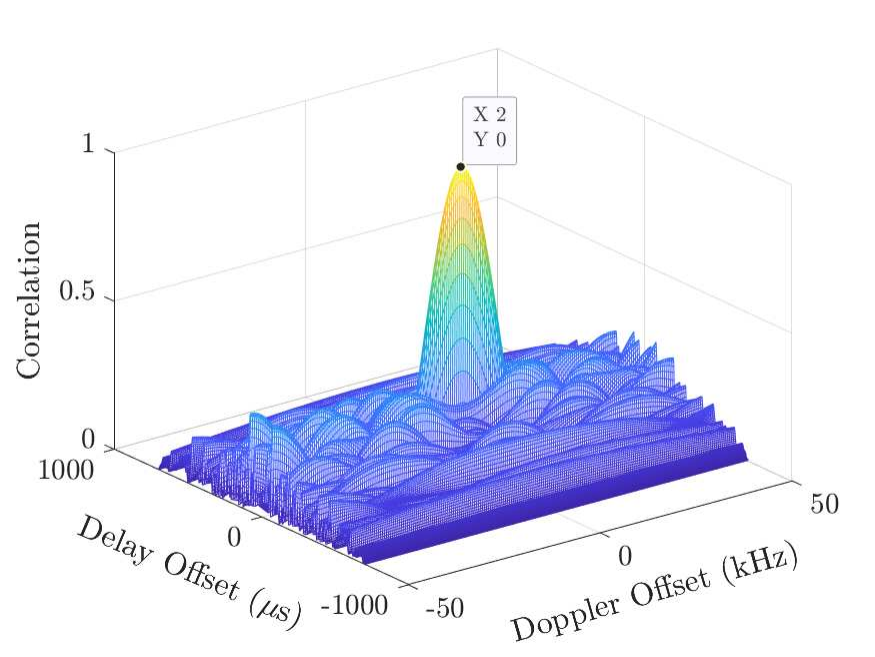}
         \captionsetup{font=small}
         \caption{Ambiguity function between \(\mathbf{Ib}_{N,0}\) and \(\mathbf{Ib}_{N,2}\) with \(-28\) kHz Doppler.}
         \label{fig:ambiguity_bN0_bN2_no_coarse}
     \end{subfigure}
     \hfill
     \centering
     \begin{subfigure}[b]{0.325\textwidth}
         \centering
         \includegraphics[width=\textwidth]{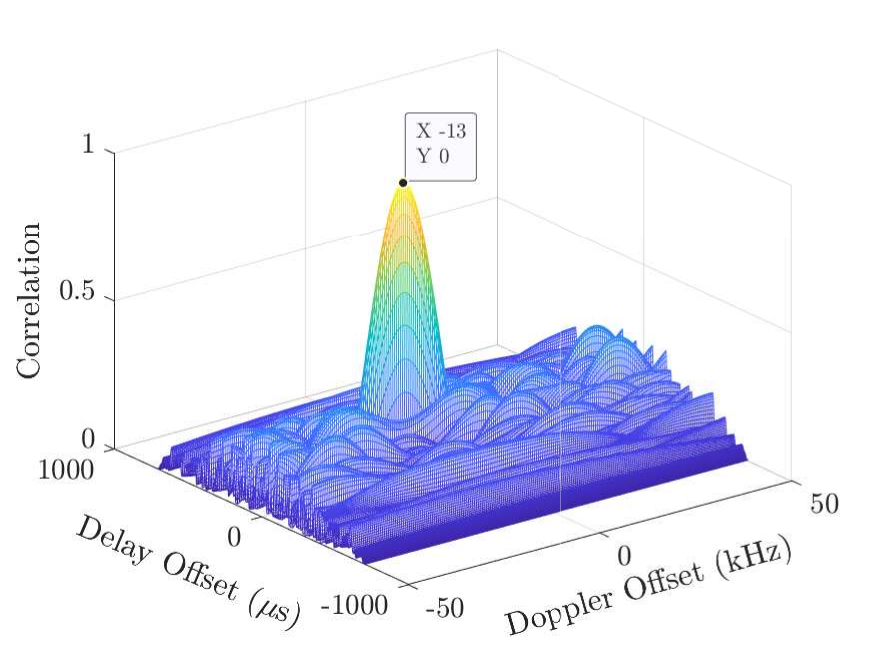}
         \captionsetup{font=small}
         \caption{Ambiguity function between \(\mathbf{Ib}_{N,1}\) and \(\mathbf{Ib}_{N,2}\) with \(-28\) kHz Doppler.}
         \label{fig:ambiguity_bN1_bN2_no_coarse}
     \end{subfigure}
      \hfill
     \begin{subfigure}[b]{0.325\textwidth}
         \centering
        \includegraphics[width=\textwidth]{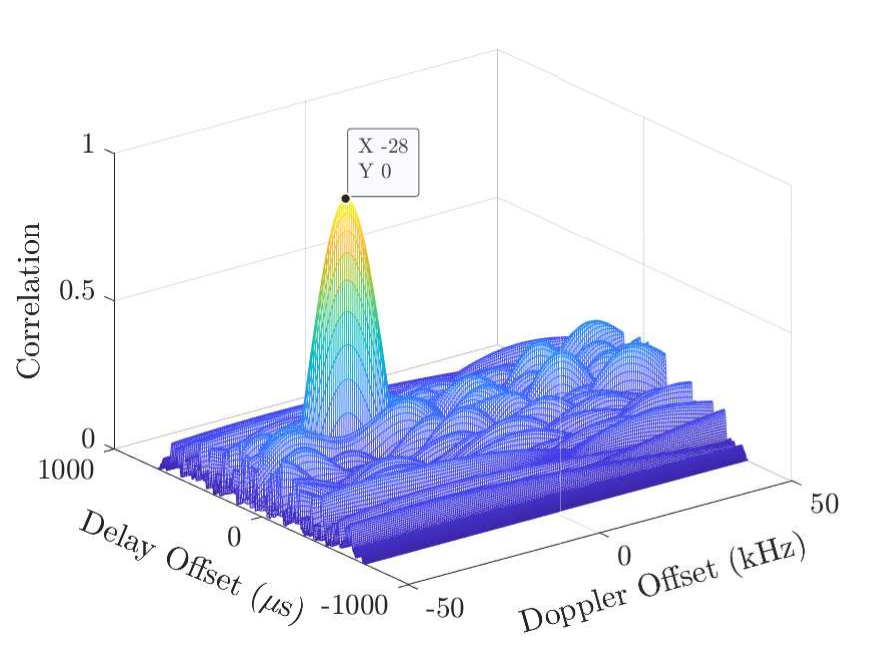}
        \captionsetup{font=small}
        \caption{Ambiguity function between \(\mathbf{Ib}_{N,2}\) and \(\mathbf{Ib}_{N,2}\) with \(-28\) kHz Doppler.}
        \label{fig:ambiguity_bN2_bN2_no_coarse}
     \end{subfigure}
     }
     \captionsetup{font=small}
     \caption{{\em Doppler-dependent behavior:} Illustrating Doppler-dependent behavior of Bj\"orck sequences.}
     \label{fig:doppler_dependent_behavior}
     \vspace{-15pt} % Adjust the negative space as needed     
\end{figure*}
\begin{figure*}[htbp]
     \centering
     \resizebox{0.90\textwidth}{!}{% Resize to match text width
     \begin{subfigure}[b]{0.325\textwidth}
         \centering
         \includegraphics[width=\textwidth]{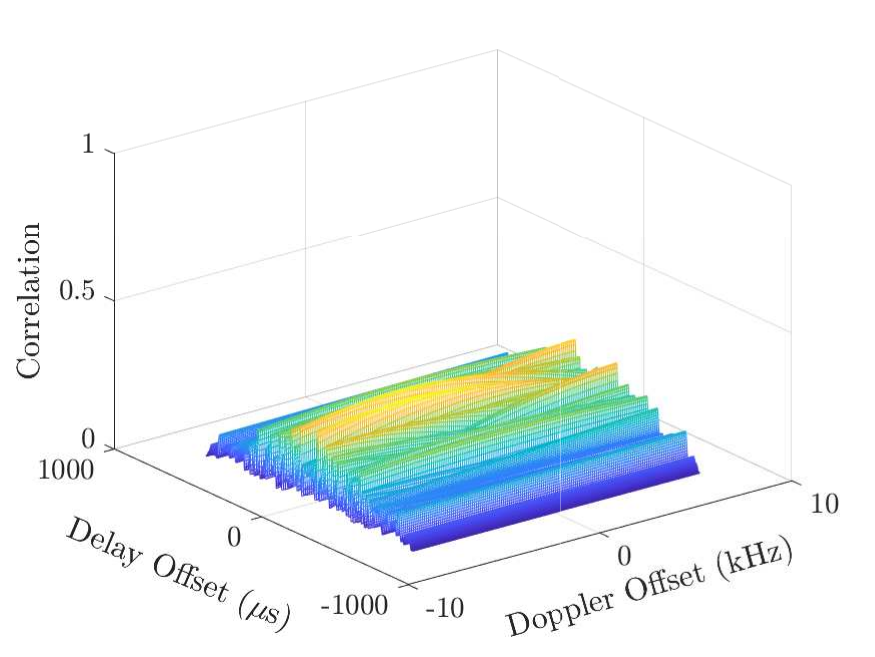}
         \captionsetup{font=small}
         \caption{Ambiguity function between \(\mathbf{Ib}_{N,0}\) and \(\mathbf{Ib}_{N,2}\) with \(-28\) kHz Doppler after correcting \(-30\) kHz.}
         \label{fig:ambiguity_bN0_bN2_coarse}
     \end{subfigure}
     \hfill
     \centering
     \begin{subfigure}[b]{0.325\textwidth}
         \centering
         \includegraphics[width=\textwidth]{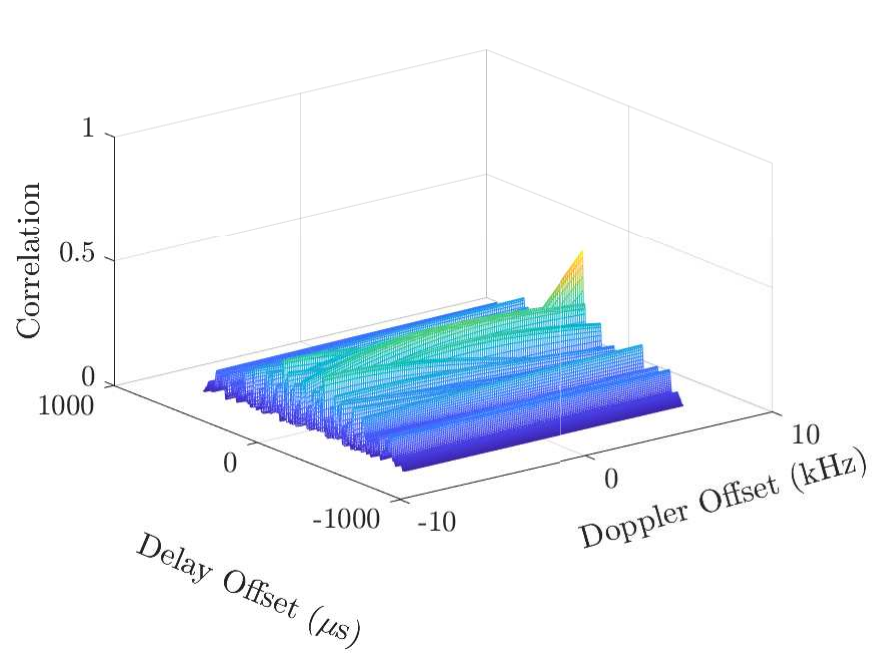}
         \captionsetup{font=small}
         \caption{Ambiguity function between \(\mathbf{Ib}_{N,1}\) and \(\mathbf{Ib}_{N,2}\) with \(-28\) kHz Doppler after correcting \(-30\) kHz.}
         \label{fig:ambiguity_bN1_bN2_coarse}
     \end{subfigure}
      \hfill
     \begin{subfigure}[b]{0.325\textwidth}
         \centering
        \includegraphics[width=\textwidth]{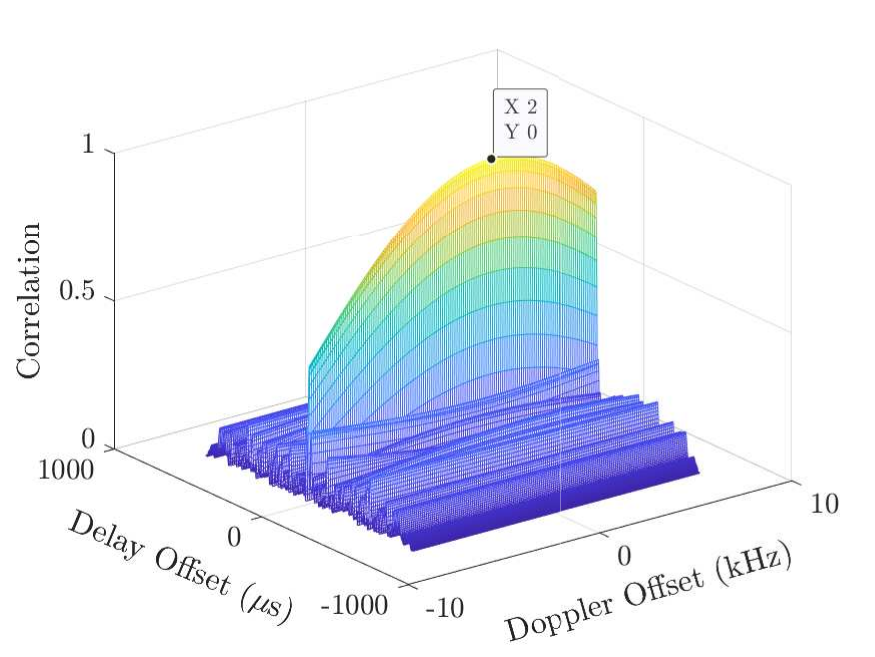}
        \captionsetup{font=small}
        \caption{Ambiguity function between \(\mathbf{Ib}_{N,2}\) and \(\mathbf{Ib}_{N,2}\) with \(-28\) kHz Doppler after correcting \(-30\) kHz.}
        \label{fig:ambiguity_bN2_bN2_coarse}
     \end{subfigure}
     }
     \captionsetup{font=small}
     \caption{{\em Approach 1: Coarse Doppler Estimate:} Using coarse Doppler estimate to address Doppler-dependent behavior of Bj\"orck sequences.}
     \label{fig:approach1_coarse_doppler}
     \vspace{-15pt} % Adjust the negative space as needed
\end{figure*}

Having established that Bj\"orck sequences exhibit superior properties and prove to be a viable alternative to the ZC sequences and a potential candidate for reference signals in wireless systems, we now address an inherent Doppler-dependent behavior. While Bj\"orck sequences demonstrate strong performance in Doppler-rich environments, this Doppler sensitivity may lead to sequence misidentification in high Doppler environments. To investigate this, we revisit the generation of orthogonal Bj\"orck sequences, where the circulant matrix \(\mathbf{B}_P\) is constructed as given in~\eqref{eq:circulant_bjorck_B^P_C}.%follows
% \begin{align}
%     \mathbf{S}_P = \begin{bmatrix}
%         \mathbf{S}_P(0) & \mathbf{S}_P(P-1) & \cdots & \mathbf{S}_P(1)\\
%         \mathbf{S}_P(1) & \mathbf{S}_P(0) & \cdots & \mathbf{S}_P(2)\\
%         \vdots & \vdots & \ddots & \vdots\\
%         \mathbf{S}_P(P-1) & \mathbf{S}_P(P-2) & \cdots & \mathbf{S}_P(0)
%     \end{bmatrix}.
% \end{align}

Let \( \mathbf{b}_{Q,l} \) represent a column of \( \mathbf{B}_Q \), where \( l \) denotes the number of cyclic rotations of the base sequence required to obtain \( \mathbf{b}_{Q,l} \). The inverse DFT (IDFT) of \( \mathbf{b}_{Q,l} \) can be expressed as
\begin{align}
        Ib_{N,l}[n] = \frac{1}{N}\sum_{m=0}^{Q-1}b_{Q,l}[m]e^{j\frac{2\pi}{N}kn},\quad n = 0,1,\dots,N-1,
\end{align}
where \( N \) is the IDFT length. We can generate \( \mathbf{b}_{Q,l} \) as
\begin{align}
    b_{Q,l}[m] = b_{Q,0}[(m-l)\mod Q)],\quad m = 0,1,\dots,Q-1.
\end{align}
where \( \mathbf{b}_{Q,0} \) is the base sequence. 

The cyclic shift in the frequency domain induces a phase ramp in the time domain, formalized in the following proposition.

\begin{prop}[Cyclic Shift and Doppler Shift Equivalence]
\label{prop:cyclic_shift_time}
Let $\mathbf{b}_{Q,l}$ denote the $l$-th cyclic shift of a base sequence in the frequency domain, and let $\mathbf{I}\mathbf{b}_{N,l}$ denote its $N$-point IDFT with samples $Ib_{N,l}[n]$. Then, the time-domain relation between $Ib_{N,0}[n]$ and $Ib_{N,l}[n]$ is given by
\begin{align}
    Ib_{N,l}[n] = Ib_{N,0}[n]e^{j\frac{2\pi}{N}nl}.
    \label{eq:zero_t_rotations}
\end{align}

This implies that \( \mathbf{Ib}_{N,l}(n) \) can be interpreted as \( \mathbf{Ib}_{N,0}(n) \) subject to a Doppler shift \( f_l \), where \( f_l \) is given by
\begin{align}
    \frac{2\pi}{N}nl = 2\pi \frac{f_l}{f_s}n, \quad f_l = l \nabla f,
\end{align}
with \( \nabla f \) denoting SCS. More generally, for any two cyclic shifts \( r \) and \( s \), the time-domain relationship extends to
\begin{align}
    Ib_{N,r}[n] = Ib_{N,s}[n]e^{j2\pi\frac{(r-s)\nabla f}{f_s}n}.
    \label{eq:s_t_rotations}    
\end{align}
\end{prop}
\begin{IEEEproof}
    This result can be proved using the frequency-shift property of the DFT.
\end{IEEEproof}

This result highlights a fundamental issue in using cyclically shifted versions of Bj\"orck sequences for high-Doppler environments, such as NTN systems, where different cyclic shifts in the frequency domain correspond to different Doppler shifts in the time domain, potentially causing ambiguity in sequence detection, thereby adversely affecting delay and Doppler measurements.

To better understand the impact of the behavior described in~\eqref{eq:zero_t_rotations} and~\eqref{eq:s_t_rotations}, consider a scenario where two satellites transmit reference signals to a specific UE. Suppose that satellite-\(1\) is assigned sequence \( \mathbf{b}_{Q,0} \) and satellite-\(2\) is assigned sequence \( \mathbf{b}_{Q,t} \). If both sequences are transmitted within the same measurement window or on the same time-frequency resources, the peak detection process may incorrectly detect satellite-\(2\) as satellite-\(1\) and attribute a Doppler shift of \(f_t\). A similar misidentification occurs if the Doppler shift between satellite-\(1\) and the UE is near zero and the Doppler shift between satellite-\(2\) and the UE is near \(-f_t\); there is a high likelihood that satellite-\(1\) could be detected as satellite-\(2\) due to identical time-domain sequences. This Doppler-dependent behavior of Bj\"orck sequences is not a significant issue in TNs, where the maximum Doppler shift is typically on the order of \(1\) kHz. To illustrate the impact of Doppler-dependent behavior, we define a variant of the ambiguity function between any two sequences \( \mathbf{Ib}_{N,s} \) and \( \mathbf{Ib}_{N,t} \). Although our subsequent analysis considers \(15\) kHz SCS for demonstration purposes, it is valid for other numerologies as well.

\begin{ndef}[A Variant of Ambiguity Function]
\label{def:gen_ambiguity_function}
Given any two sequences \( \mathbf{Ib}_{N,s} \) and \( \mathbf{Ib}_{N,t} \), we define a variant of the discrete ambiguity function as:
\begin{align}
    \mathbf{A}_{t,s}(n, k) = \frac{1}{N}\sum_{l=0}^{N-1} \mathbf{Ib}_{N,s}(n+l)\mathbf{Ib}_{N,t}^*(l)e^{-j2\pi\frac{f_k}{f_s}l}.
    \label{eq:gen_ambiguity_fn}
\end{align}
where \( n \) represents the delay shift, \( k \) denotes the Doppler shift hypothesis, and \( f_s \) is the sampling frequency.
\end{ndef}

Consider a scenario where \( \mathbf{Ib}_{N,2} \), generated assuming \(15\) kHz SCS, is assigned to a specific LEO-UE link with a Doppler shift of \(-28\) kHz. Fig.~\ref{fig:ambiguity_bN0_bN2_no_coarse}, Fig.~\ref{fig:ambiguity_bN1_bN2_no_coarse}, and Fig.~\ref{fig:ambiguity_bN2_bN2_no_coarse} show the ambiguity functions plotted by correlating \( \mathbf{Ib}_{N,2} \), subject to a Doppler shift of \(-28\) kHz, with locally generated sequences \( \mathbf{Ib}_{N,0} \), \( \mathbf{Ib}_{N,1} \), and \( \mathbf{Ib}_{N,2} \), respectively. It is evident that due to the Doppler-dependent behavior described in~\eqref{eq:zero_t_rotations} and~\eqref{eq:s_t_rotations}, the sequence \( \mathbf{Ib}_{N,2} \) with a \(-28\) kHz Doppler shift can be misidentified as \( \mathbf{Ib}_{N,0} \) with a \(2\) kHz Doppler shift or \( \mathbf{Ib}_{N,1} \) with a \(-13\) kHz Doppler shift. 

\subsection{Mitigation Approaches}\label{sec:doppler_behavior_mitigation}
To address misidentification due to Doppler-dependent behavior, we propose two approaches: 1) assuming the availability of a coarse Doppler estimate and 2) using a subset of available sequences such that the frequency separation among the sequences exceeds the maximum Doppler uncertainty. 

{\em Approach 1: Coarse Doppler Estimate. }In the first approach, we assume the availability of a coarse Doppler estimate. The system first compensates for this estimate and then performs peak detection over a narrowed time-frequency search space. This approach requires that the coarse Doppler estimate does not deviate from the actual Doppler by more than half of the SCS. Specifically, the desired peak must lie within the range \([-0.5 \nabla f, \,0.5 \nabla f]\). For instance, in the case of a \(15\) kHz SCS, the deviation should not exceed \(7.5\) kHz. To evaluate this approach, consider the previous scenario where \( \mathbf{Ib}_{N,2} \) is subject to a Doppler shift of \(-28\) kHz, considering \(15\) kHz SCS. Assuming a coarse Doppler estimate of \(-30\) kHz, we perform peak detection after compensating for this estimate and narrowing the Doppler hypothesis to the range \([-7.5,\,7.5]\) kHz. Fig.~\ref{fig:ambiguity_bN0_bN2_coarse}, Fig.~\ref{fig:ambiguity_bN1_bN2_coarse}, and Fig.~\ref{fig:ambiguity_bN2_bN2_coarse} illustrate the corresponding ambiguity functions for sequences \( \mathbf{Ib}_{N,0} \), \( \mathbf{Ib}_{N,1} \), and \( \mathbf{Ib}_{N,2} \), respectively. With the coarse Doppler compensation and narrowed search space, \( \mathbf{Ib}_{N,2} \) exhibits a distinct peak, effectively avoiding misidentification. This distinct peak facilitates the estimation of both the delay and the residual frequency offset. Notably, the peak appears broader due to the narrower Doppler hypothesis range. However, the downside of this approach is that other reference signals, such as the synchronization signal block (SSB)~\cite{lin2021dopplershiftestimation5g}, are required to obtain the coarse Doppler estimate.
\begin{figure*}[htbp]
     \centering
     \resizebox{0.90\textwidth}{!}{% Resize to match text width
     \begin{subfigure}[b]{0.325\textwidth}
         \centering
         \includegraphics[width=\textwidth]{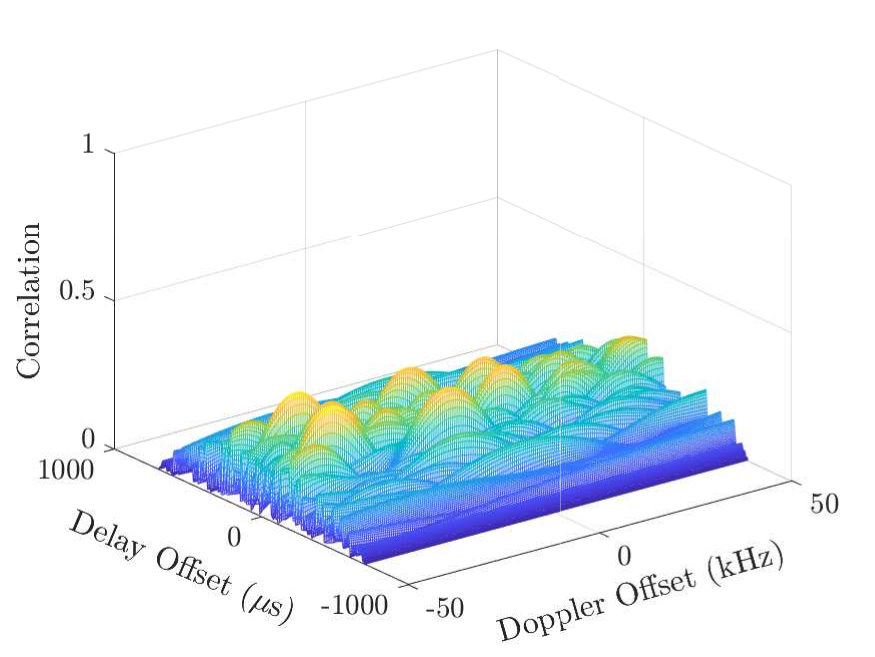}
         \captionsetup{font=small}
         \caption{Ambiguity function between \(\mathbf{Ib}_{N,0}\) and \(\mathbf{Ib}_{N,7}\) with \(-42\) kHz Doppler.}
         \label{fig:ambiguity_bN0_bN7}
     \end{subfigure}
     \hfill
     \centering
     \begin{subfigure}[b]{0.325\textwidth}
         \centering
         \includegraphics[width=\textwidth]{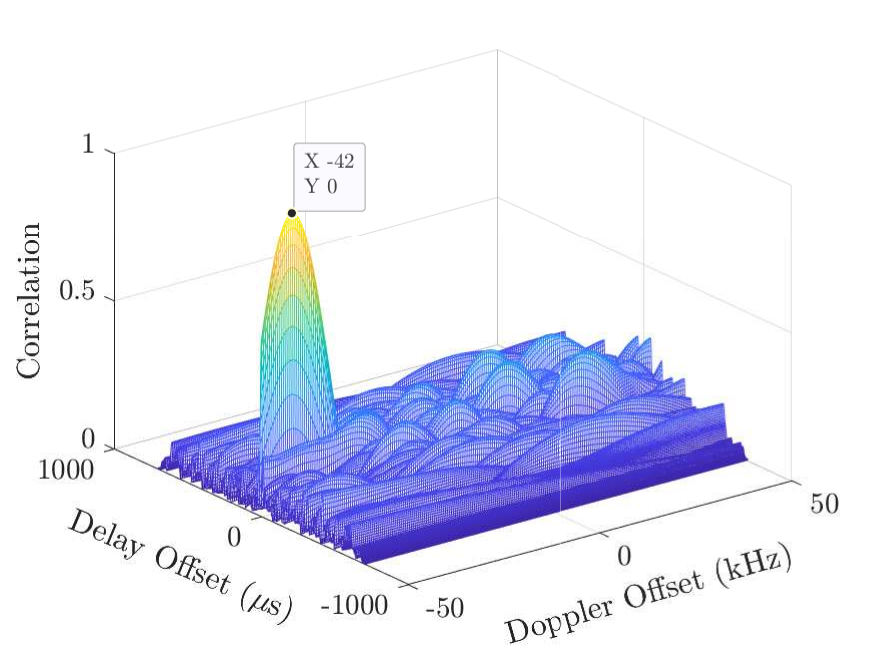}
         \captionsetup{font=small}
         \caption{Ambiguity function between \(\mathbf{Ib}_{N,7}\) and \(\mathbf{Ib}_{N,7}\) with \(-42\) kHz Doppler.}
         \label{fig:ambiguity_bN7_bN7}
     \end{subfigure}
     \hfill
     \centering
     \begin{subfigure}[b]{0.325\textwidth}
         \centering
         \includegraphics[width=\textwidth]{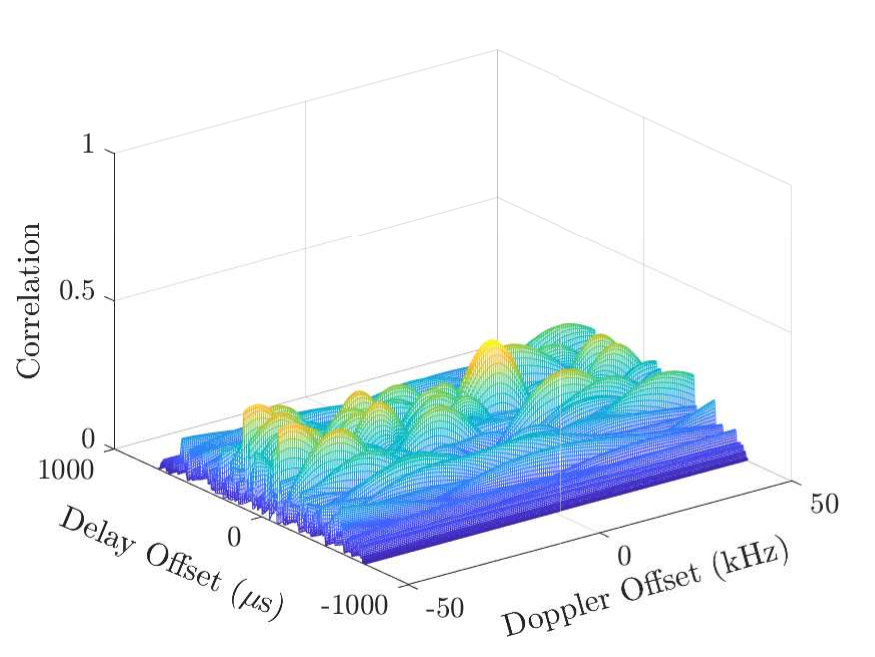}
         \captionsetup{font=small}
         \caption{Ambiguity function between \(\mathbf{Ib}_{N,14}\) and \(\mathbf{Ib}_{N,7}\) with \(-42\) kHz Doppler.}
         \label{fig:ambiguity_bN14_bN7}
     \end{subfigure}
     }
     \captionsetup{font=small}
     \caption{{\em Approach 2: Sequence Subset Selection:} Using a subset of available sequences to maintain a minimum separation of more than double the Doppler shift.}
     \label{fig:approach2_seq_subset}
     \vspace{-15pt} % Adjust the negative space as needed
\end{figure*}

{\em Approach 2: Sequence Subset Selection. }The second approach proposes selecting a subset of available orthogonal sequences based on the maximum Doppler shifts observed in the system. The Doppler shift can vary significantly with the elevation angle in LEO-based systems. %,emenonye2024joint9dreceiverlocalization,emenonye2024minimal,emenonye2024unsync}. 
For example, the difference in Doppler shifts between an elevation angle of \(90^\circ\) and as low as \(10^\circ\) can be as large as \(45\) kHz~\cite{dureppagari2024leo}. Given \(15\) kHz SCS, this corresponds to approximately three frequency shifts. A separation of more than double this value, i.e., seven frequency shifts, must be maintained between the sequences assigned to different LEOs to prevent misidentification. This ensures that the sequences remain distinguishable despite high Doppler shifts. To evaluate this approach, consider a scenario where the sequence \( \mathbf{Ib}_{N,7} \), generated using a \(1\) MHz bandwidth and \(15\) kHz SCS, is assigned to a specific LEO-UE link experiencing a Doppler shift of \(-42\) kHz. Fig.~\ref{fig:ambiguity_bN0_bN7}, Fig.~\ref{fig:ambiguity_bN7_bN7}, and Fig.~\ref{fig:ambiguity_bN14_bN7} illustrate the ambiguity functions obtained by correlating \( \mathbf{Ib}_{N,7} \), subject to a Doppler shift of \(-42\) kHz, with locally generated sequences \( \mathbf{Ib}_{N,0} \), \( \mathbf{Ib}_{N,7} \), and \( \mathbf{Ib}_{N,14} \), respectively. By maintaining a separation of more than seven frequency shifts, it is evident that only \( \mathbf{Ib}_{N,7} \) produces a clear peak, successfully avoiding misidentification. 

However, approach 2 has a notable drawback: it reduces the number of sequences available for assignment. For NTN systems, especially with large constellations and applications such as positioning~\cite{dureppagari_ntn_10355106,dureppagari2024leo,dureppagari2026leobasedcarrierphasepositioning6g} that require multiple satellites communicating with a single UE, the limitation could pose a significant challenge. To address this limitation, sequence reuse can be enabled by partitioning the coverage into NTN reuse cells and assigning sequences either uniquely to each cell or reused across spatially separated cells (e.g., using a reuse factor \(K\) in a hexagonal layout), analogous to similar to frequency reuse \(1\) and \(K\). While the former ensures minimal interference, the latter improves resource efficiency by allowing reuse among non-adjacent regions at the cost of reduced sequence diversity. A detailed investigation of sequence reuse strategies is beyond the scope of this paper and is dealt with in future work.

\section{Concluding Remarks}\label{sec:conclusion}

In this paper, we developed a generic construction framework for extending prime-length Bj\"orck sequences to arbitrary lengths by leveraging Goldbach's conjecture. The proposed approach is general and applies to any CAZAC family inherently defined over prime lengths. A key contribution of this work is the systematic characterization of the impact of such extensions on correlation properties. In particular, we showed that, for cyclically shifted constructions, perfect zero-lag orthogonality can be preserved under strict non-overlapping conditions, which inherently limits the number of mutually orthogonal sequences to $\min\{Q_1, Q_2\}$. This reveals a fundamental tradeoff between the total number of sequences and the number of orthogonal sequences, determined by the choice of $(Q_1,\,Q_2)$. In contrast, for extending different root index sequences, we demonstrated that the normalized inner product remains inherently bounded and scales on the order of $1/\sqrt{N}$, thereby eliminating the need for strict non-overlapping constraints while maintaining controlled interference levels. Beyond zero-lag analysis, we provided a detailed study of periodic and aperiodic cross-correlation properties using RMS-based metrics. 
% Our analysis highlights a fundamental distinction between terrestrial and non-terrestrial network regimes. In terrestrial networks, where signals from different transmitters arrive with negligible delay and Doppler differences, periodic cross-correlation is the dominant metric, and orthogonality can be effectively exploited through cyclic shifts. However, in LEO-based NTN systems, large propagation delays and significant Doppler variations result in partial overlap and misalignment of reference signals at the receiver. In such scenarios, aperiodic cross-correlation becomes the more relevant metric. 
Importantly, we showed that the proposed construction preserves the favorable $1/\sqrt{N}$ scaling behavior for both periodic and aperiodic RMS correlation, with aperiodic correlation often benefiting from reduced effective overlap. Building on this framework, we investigated Bj\"orck sequences as a potential candidate for reference signals in wireless systems for both TN and NTN systems. In particular, their superior ambiguity function enables improved delay and Doppler estimation performance compared to ZC sequences in NTN scenarios. At the same time, we identified an inherent Doppler-dependent behavior that may lead to sequence misidentification under large Doppler shifts. To address this, we proposed two practical mitigation strategies: (i) leveraging coarse Doppler estimation prior to detection and (ii) selecting appropriately spaced subsets of sequences to account for maximum Doppler uncertainty. Overall, the proposed sequence construction framework, together with the demonstrated properties of Bj\"orck sequences, provides a unified and flexible solution for reference signal design in next-generation wireless systems.

\appendix

\subsection{Proof of Lemma~\ref{lem:per_rms_extended_cyclic}}\label{app::periodic_cross_corr}
Define the unnormalized periodic cross-correlation
\begin{align}
R_{ij}(\tau)
=
\sum_{n=0}^{N-1} s_i[n]\;s_j^*[(n-\tau)\bmod N],\,
C_{ij}(\tau)=\frac{1}{N}R_{ij}(\tau).
\label{eq:lemma_per_def_Rij}
\end{align}

For \(0\le \tau \le Q_1\), the circular shift induces four block-overlap terms, yielding the exact decomposition
\begin{align}
R_{ij}(\tau)
&=
\underbrace{\sum_{n=\tau}^{Q_1-1} x_i[n]\;x_j^*[n-\tau]}_{\text{top--top}}
+
\underbrace{\sum_{n=0}^{\tau-1} x_i[n]\;y_{\pi(j)}^*[(n-\tau)\bmod Q_2]}_{\text{top--bottom}}
\notag\\
&\quad+
\underbrace{\sum_{m=0}^{\tau-1} y_{\pi(i)}[m]\;x_j^*[(Q_1-\tau+m)\bmod Q_1]}_{\text{bottom--top}}\notag\\
&\quad+\underbrace{\sum_{m=\tau}^{Q_2-1} y_{\pi(i)}[m]\;y_{\pi(j)}^*[m-\tau]}_{\text{bottom--bottom}},
\label{eq:lemma_per_exact_block_decomp}
\end{align}
where indices of \(x_j[\cdot]\) are interpreted modulo \(Q_1\) and indices of \(y_{\pi(j)}[\cdot]\) modulo \(Q_2\). For \(Q_1<\tau\le N-1\), the remaining lags follow from circular-correlation symmetry as
\begin{align}
R_{ij}(\tau)=R_{ji}^*(N-\tau).
\label{eq:lemma_per_symmetry}
\end{align}

Notably, by setting \(\tau=0\) in~\eqref{eq:lemma_per_exact_block_decomp}, we obtain
\begin{align}
C_{ij}(0)
=
\frac{1}{N}R_{ij}(0)
=
\begin{cases}
0, & \pi(i)\neq \pi(j),\\[4pt]
\dfrac{Q_2}{N}, & \pi(i)=\pi(j).
\end{cases}
\label{eq:periodic_zero_lag_exps}
\end{align}
which are essentially the inner-product terms derived in Lemma~\ref{lem:inner_product_even_cyclic}. 

Next, from~\eqref{eq:lemma_per_rms_def} and~\eqref{eq:lemma_per_def_Rij},
\begin{align}
\mathrm{RMS}_{ij}^2
=
\frac{1}{N}\sum_{\tau=0}^{N-1}\left|\frac{R_{ij}(\tau)}{N}\right|^2
=
\frac{1}{N^3}\sum_{\tau=0}^{N-1}|R_{ij}(\tau)|^2.
\label{eq:lemma_per_rms_exact_identity}
\end{align}

% To obtain closed-form expressions for~\eqref{eq:lemma_per_rms_exact_identity}, we use the standard random-phase approximation for nonzero lags, under which each term in \eqref{eq:lemma_per_exact_block_decomp} behaves like a sum of approximately random unit-modulus phasors. Consequently, for \(\tau\neq 0\),
% \begin{align}
% \mathbb{E}|R_{ij}(\tau)|^2 \approx N,
% \qquad
% \mathbb{E}|C_{ij}(\tau)|^2 \approx \frac{1}{N}.
% \label{eq:lemma_per_nonzero_scaling}
% \end{align}
To obtain closed-form expressions for~\eqref{eq:lemma_per_rms_exact_identity}, we use a standard approximation for sums of unit-modulus phasors~\cite{gregoratti2023mathematicalpropertieszadoffchusequences}. Specifically, consider a sum of the form
\begin{align}
S_L = \sum_{\ell=1}^{L} z_\ell, \qquad |z_\ell|=1,
\end{align}
where the phases of \(z_\ell = e^{j\theta_\ell}\) are approximately independent and uniformly distributed over \([0,2\pi)\). Then, it is known that
\begin{align}
\mathbb{E}\!\left[|S_L|^2\right]
=
\sum_{\ell=1}^{L} \mathbb{E}[|z_\ell|^2]
+
\sum_{\ell\neq k} \mathbb{E}[z_\ell z_k^*]
\approx L,
\label{eq:rand_phase_identity}
\end{align}
since the cross terms vanish due to phase incoherence.

This approximation is applicable to \eqref{eq:lemma_per_exact_block_decomp} for \(\tau\neq 0\), because each term in the four overlap components corresponds to a product of CAZAC sequence elements with unit magnitude but varying phase. For nonzero lags, the perfect phase alignment present at \(\tau=0\) is destroyed, and the resulting phasors behave approximately as independent random variables.

In particular, for \(0 \le \tau \le Q_2\), the four overlap terms in \eqref{eq:lemma_per_exact_block_decomp} have lengths
\begin{align}
L_{xx} = Q_1 - \tau, \quad
L_{xy} = \tau, \quad
L_{yx} = \tau, \quad
L_{yy} = Q_2 - \tau.
\end{align}

Applying \eqref{eq:rand_phase_identity} to each component at \(\tau\neq 0\) and summing the contributions, we obtain
\begin{align}
\mathbb{E}|R_{ij}(\tau)|^2
&\approx
L_{xx} + L_{xy} + L_{yx} + L_{yy} \notag\\
&=
(Q_1-\tau) + \tau + \tau + (Q_2-\tau)
= Q_1 + Q_2
= N.
\label{eq:rand_phase_periodic_sum}
\end{align}

For \(Q_2 < \tau \le Q_1\), the bottom--bottom term vanishes, but the remaining overlap lengths sum to
\begin{align}
(Q_1-\tau) + \tau + \tau = Q_1 + \tau \le N,
\end{align}
which remains on the order of \(N\). Similarly, for \(Q_1 < \tau \le N-1\), the result follows from symmetry in~\eqref{eq:lemma_per_symmetry}.

Therefore, for all nonzero lags, i.e., \(\tau\neq0\),
\begin{align}
\mathbb{E}|R_{ij}(\tau)|^2 \approx N,
\qquad
\mathbb{E}|C_{ij}(\tau)|^2 \approx \frac{1}{N}.
\label{eq:lemma_per_nonzero_scaling}
\end{align}

We now evaluate the two cases.

\textit{Case 1: \(\pi(i)\neq \pi(j)\).}
Using \(C_{ij}(0)=0\) and \eqref{eq:lemma_per_nonzero_scaling}, we get
\begin{align}
\mathrm{RMS}_{ij}^2
&\approx
\frac{1}{N}\sum_{\tau=1}^{N-1}\frac{1}{N}
=
\frac{N-1}{N^2},
\end{align}
which gives
\begin{align}
\mathrm{RMS}_{ij}
\approx
\frac{1}{\sqrt{N}}\sqrt{1-\frac{1}{N}}.
\end{align}
% This proves \eqref{eq:lemma_per_rms_case1}.

\textit{Case 2: \(\pi(i)=\pi(j)\).}
Using \(C_{ij}(0)=Q_2/N\) and \eqref{eq:lemma_per_nonzero_scaling},
\begin{align}
\mathrm{RMS}_{ij}^2
&\approx
\frac{1}{N}\left[
\left(\frac{Q_2}{N}\right)^2
+
(N-1)\frac{1}{N}
\right]
=
\frac{Q_2^2}{N^3}+\frac{N-1}{N^2},
\end{align}
and therefore
\begin{align}
\mathrm{RMS}_{ij}
\approx
\frac{1}{\sqrt{N}}
\sqrt{1-\frac{1}{N}+\left(\frac{Q_2}{N}\right)^2}.
\end{align}
% This proves \eqref{eq:lemma_per_rms_case2}. Equation \eqref{eq:lemma_per_rms_ratio} follows immediately by normalization with respect to \(1/\sqrt{N}\).

\subsection{Proof of Lemma~\ref{lem:aper_rms_extended_cyclic}}\label{app::aperiodic_cross_corr}
For \(0\le \tau\le N-1\), the unnormalized aperiodic cross-correlation is
\begin{align}
R_{ij}^{(a)}(\tau)
=
\sum_{n=0}^{N-1-\tau} s_i[n+\tau]s_j^*[n].
\label{eq:aper_proof_start}
\end{align}

% Since no circular wrap occurs, the correlation window has length \(N-\tau\). 
For \(0\le \tau\le Q_1\), the overlap splits into three terms:
\begin{align}
R_{ij}^{(a)}(\tau)
&=
\underbrace{\sum_{n=0}^{Q_1-1-\tau}
x_i[n+\tau]x_j^*[n]}_{\text{top--top, length }Q_1-\tau}
+
\underbrace{\sum_{n=Q_1-\tau}^{Q_1-1}
y_{\pi(i)}[n+\tau-Q_1]x_j^*[n]}_{\text{bottom--top, length }\tau}
\notag\\
&\quad+
\underbrace{\sum_{n=Q_1}^{N-1-\tau}
y_{\pi(i)}[n+\tau-Q_1]y_{\pi(j)}^*[n-Q_1]}_{\text{bottom--bottom, length }Q_2-\tau\ \text{if }\tau\le Q_2}.
\label{eq:aper_proof_decomp}
\end{align}

For negative lags, the standard symmetry
\begin{align}
R_{ij}^{(a)}(-\tau)=\big(R_{ji}^{(a)}(\tau)\big)^*
\end{align}
holds.

At zero lag, similar to the periodic case, we have
\begin{align}
C_{ij}^{(a)}(0)
=
\frac{1}{N}R_{ij}^{(a)}(0)
=
\begin{cases}
0, & \pi(i)\neq \pi(j),\\[4pt]
\dfrac{Q_2}{N}, & \pi(i)=\pi(j).
\end{cases}
\label{eq:aper_proof_tau0_final}
\end{align}

Next, from \eqref{eq:aper_lemma_rms} and the symmetry of aperiodic correlation,
\begin{align}
\mathrm{MS}_{ij}^{(a)}
=
\frac{1}{2N-1}
\left(
\frac{|R_{ij}^{(a)}(0)|^2}{N^2}
+
\frac{2}{N^2}
\sum_{\tau=1}^{N-1}|R_{ij}^{(a)}(\tau)|^2
\right).
\label{eq:aper_proof_ms}
\end{align}

To approximate the nonzero-lag terms, we use the standard random-phase model for sums of unit-modulus phasors
\begin{align}
\mathbb{E}\left|\sum_{\ell=1}^{L} z_\ell \right|^2 \approx L.
\label{eq:aper_rand_phase}
\end{align}

Under this approximation, for \(1\le \tau\le Q_2\), all three terms in \eqref{eq:aper_proof_decomp} are present, and the total overlap length is
\begin{align}
(Q_1-\tau)+\tau+(Q_2-\tau)=N-\tau,
\end{align}
so that
\begin{align}
\mathbb{E}|R_{ij}^{(a)}(\tau)|^2 \approx N-\tau.
\label{eq:aper_proof_reg1}
\end{align}

For \(Q_2<\tau\le Q_1-1\), the bottom--bottom term disappears, and the total overlap becomes
\begin{align}
(Q_1-\tau)+\tau=Q_1,
\end{align}
hence
\begin{align}
\mathbb{E}|R_{ij}^{(a)}(\tau)|^2 \approx Q_1.
\label{eq:aper_proof_reg2}
\end{align}

For \(Q_1\le \tau\le N-1\), only a tail overlap remains, whose length is \(N-\tau\), yielding
\begin{align}
\mathbb{E}|R_{ij}^{(a)}(\tau)|^2 \approx N-\tau.
\label{eq:aper_proof_reg3}
\end{align}

Summing \eqref{eq:aper_proof_reg1}--\eqref{eq:aper_proof_reg3} over all positive lags gives
\begin{align}
\sum_{\tau=1}^{N-1}\mathbb{E}|R_{ij}^{(a)}(\tau)|^2
\approx
Q_2N+Q_1(Q_1-Q_2-1)
=P.
\label{eq:aper_proof_P}
\end{align}

Let \(P = Q_2N+Q_1(Q_1-Q_2-1)\). Substituting \(P\) into~\eqref{eq:aper_proof_ms} yields
\begin{align}
\mathbb{E}\!\left[\mathrm{MS}_{ij}^{(a)}\right]
\approx
\frac{1}{2N-1}
\left(
\frac{|R_{ij}^{(a)}(0)|^2}{N^2}
+
\frac{2P}{N^2}
\right).
\label{eq:aper_proof_final_ms}
\end{align}

If \(\pi(i)\neq \pi(j)\), then \(R_{ij}^{(a)}(0)=0\), and \eqref{eq:aper_case1_ms} follows immediately.

If \(\pi(i)=\pi(j)\), then \(R_{ij}^{(a)}(0)=Q_2\), and \eqref{eq:aper_case2_ms} follows.

Finally, since the dominant term in \(P\) scales as \(Q_1^2\), the RMS behaves as
\begin{align}
\mathbb{E}\!\left[\mathrm{RMS}_{ij}^{(a)}\right]
\approx
\sqrt{\frac{P}{N^3}}
\approx
\frac{Q_1}{N}\frac{1}{\sqrt{N}}
=
\left(1-\frac{Q_2}{N}\right)\frac{1}{\sqrt{N}}.
\end{align}

\bibliographystyle{IEEEtran}
\bibliography{hokie}
\end{document}